\documentclass[preprint,12pt,authoryear]{elsarticle}

\usepackage{amssymb}
\usepackage{natbib}
\usepackage{algorithm}
\usepackage{amsmath}
\usepackage{mathrsfs}
\usepackage[margin=1.0in]{geometry}
\usepackage{graphicx}
\usepackage{subcaption}

\journal{arXiv}

\begin{document}
\begin{frontmatter}

\title{Bayesian Computation for Log-Gaussian Cox Processes--A Comparative Analysis of Methods}

\author{Ming Teng\fnref{umich}}  \author{Timothy D. Johnson\fnref{umich}} \author{Farouk S. Nathoo\corref{cor1}\fnref{uvic}} 

\fntext[umich]{University of Michigan, Department of Biostatistics, 1415 Washington Heights, Ann Arbor, MI 48109}
\fntext[uvic]{University of Victoria, Department of Mathematics and Statistics, Victoria, BC, Canada, V8W 3P4}
\cortext[cor1]{Corresponding author. Tel: +1 250 472 4693. Email: nathoo@uvic.ca}

\begin{abstract}
The Log-Gaussian Cox Process is a commonly used model for the analysis of spatial point patterns. Fitting this model is difficult because of its doubly-stochastic property, i.e., it is an hierarchical combination of a Poisson process at the first level and a Gaussian Process at the second level.  Different methods have been proposed to estimate such a process, including traditional likelihood-based approaches as well as Bayesian methods. We focus here on Bayesian methods and several approaches that have been considered for model fitting within this framework, including Hamiltonian Monte Carlo, the Integrated nested Laplace approximation, and Variational Bayes. We consider these approaches and make comparisons with respect to statistical and computational efficiency. These comparisons are made through several simulations studies as well as through applications examining both ecological data and neuroimaging data. 

\end{abstract}


\end{frontmatter}

\section{Introduction}

Spatial point process models for point pattern data have many applications including those involving ecology, geology, seismology, and neuroimaging. The theory of point processes has its roots with the work of Poisson in the 19th century, while more modern treatments with a statistical focus include \cite{daley1988introduction}, \cite{moller1998log} and \cite{illian2008statistical}. Among models for a spatial point process, the homogeneous Poisson process is the most fundamental but its use is limited in many applications due to its simplistic nature. A related but more flexible process is the Log-Gaussian Cox Process (LGCP), a process that is obtained by assuming a hierarchical structure, where at the first level the process is assumed Poisson conditional on the intensity function, and at the second level the log of the intensity function is assumed to be drawn from a Gaussian process. The flexibility of the model arises from the Gaussian process prior specified over the log-intensity function. Given this hierarchical structure with a Gaussian process at the second level, fitting this model to observed spatial point pattern data is a computational challenge \citep{murray2012mcmc}. 

A number of approaches have been developed to estimate the LGCP model in both the classical and Bayesian frameworks. In \cite{diggle1985kernel} the authors propose an adaption of Rosenblatt's kernel method \citep{rosenblatt1956remarks} for the purpose of non-parametric estimation and then derive an expression for the mean squared error based on a stationarity assumption. A Bayesian framework is considered in \cite{moller1998log} where the authors propose the use of the Metropolis-Adjusted Langevin Algorithm (MALA) \citep{besag1994discussion} for Monte Carlo sampling of the posterior distribution. These authors also introduce a discretization of the spatial domain in order to attain computational tractability. \cite{adams2009tractable} proposes an exact estimation method to deal with a modification of such a point process which they term the Sigmoidal Gaussian Cox process. MALA is a special case of the potentially more efficient Hamiltonian Monte Carlo (HMC) \citep{neal1995bayesian} algorithm that uses the notion of Hamiltonian dynamics to construct proposals for the Metropolis-Hastings algorithm and has been adopted recently for an increasing number of applications \citep{neal2011mcmc}. Extending further, the Riemann Manifold Hamiltonian Monte Carlo algorithm proposed by \cite{girolami2011riemann} is a generalization of both MALA and HMC that can lead to more efficient posterior sampling in some cases. One drawback of Riemann Manifold HMC is that it requires the inversion of a potentially large matrix (the expected Fisher information matrix plus the negative Hessian of the log prior) at each iteration and this is not computationally feasible for very high-dimensional problems such as those typically involving the LGCP model.

An advantage associated with the use of MCMC algorithms for Bayesian computation is the underlying theory which guarantees simulation consistent estimation of various important characteristics of the posterior distribution. Thus the practitioner is assured of an accurate Monte Carlo representation of the posterior distribution given a sufficient amount of sampling effort. A drawback is that MCMC can be computationally intense, and this has motivated several alternative deterministic approaches for approximate Bayesian inference. 

One such approach is Variational Bayes (VB) \citep{mackay1997ensemble}, where the approximation to the posterior distribution is assumed to lie within some convenient family of distributions and then an optimization is carried out to minimize the Kullback-Leibler divergence measuring the discrepancy between the true posterior and the approximation.  Often the family of distributions within which the approximation is assumed to lie is based on the notion of a mean field approximation, which corresponds to assuming posterior independence between certain model parameters. The idea in this case is to replace stochastic posterior dependence between parameters with deterministic dependence between the posterior moments in a manner that minimizes Kullback-Leibler divergence. Variational Bayes approximations have been applied successfully to the analysis of hidden Markov models in \cite{mackay1997ensemble} and to other mixture models \cite{humphreys2000approximate}. \cite{zammit2012variational} used Variational Bayes for models of spatiotemporal systems represented by linear stochastic differential equations and demonstrated quick and efficient approximate inference both for continuous observations and point process data. 

Mean field Variational Bayes is well suited for dealing with models within the conjugate exponential family where closed form solutions for the iterative steps of the optimization algorithm are available. In general such closed form solutions may not be available and additional approximations are then required. This is the case with the LGCP model. Mean field Variational Bayes approximations for non-conjugate models can be obtained by incorporating further approximations based on the delta method or the Laplace method  \cite{wang2013variational}. These approximations are successfully applied to a correlated topic model and a Bayesian logistic regression model in \cite{wang2013variational}. For the LGCP model, tractable variational approximations can be obtained following this approach, where a mean field approximation with further approximations based on the Laplace method are used to handle the non-conjugate structure of the model. 

Variational Bayes approximations can work well in some settings and the corresponding approximations can be computed relatively fast. A drawback is that there is no underlying theory guaranteeing the accuracy of the approximation or characterizing its error, thus these approximations need to be evaluated on a case-by-case basis, and they may or may not achieve reasonable accuracy depending on the utility of the practitioner. One contribution of this paper is the derivation of a mean field VB approximation which incorporates the Laplace method for the LGCP model. As far as we are aware such approximations have not been considered previously for this model. Another contribution of this paper is to compare this VB approximation  with HMC in terms of both statistical and computational efficiency. 

An alternative approach for approximate Bayesian inference that has gained tremendous popularity in the statistical literature is the integrated nested Laplace approximation (INLA) \citep{rue2009approximate}. INLA is less generally applicable than MCMC or VB as it assumes the model has a latent Gaussian structure with only a moderate number of hyper-parameters. For spatial modeling the approach makes use of the Gaussian Markov Random field \citep{rue2005gaussian} and corresponding approximations which are known to be computationally efficient. The basis of INLA is the use of the Laplace approximation and numerical integration with latent Gaussian models to derive approximate posterior marginal distributions. INLA does not produce an approximation to the joint posterior which is a drawback of the approach in settings where the joint posterior (as opposed to the marginals) is of interest. 

For spatial models incorporating a Gaussian Random field (GRF) with a Mat\'{e}rn correlation structure, \cite{lindgren2011explicit} develop an approximate approach based on stochastic partial differential equations (SPDE) and these approximations have been combined for use with INLA. The essence of the approach is to specify a SPDE that has as its solution the GRF and then the SPDE representation is used in conjunction with basis representations to approximate the process over the vertices of a 2-dimensional mesh covering the spatial domain. The value of the process at any location is then obtained based on interpolation of the values at the mesh vertices. In recent work, \cite{simpson2012think} evaluated this approximation applied to the LGCP model with spatially varying covariates and demonstrated adequate performance for the settings and data considered there. 

A comparison between INLA and MALA for models incorporating GMRF approximations is considered in \cite{taylor2013inla}. In our work, we compare for the LGCP model Bayesian computation based on HMC, VB with a Laplace approximation, INLA,  and INLA with the SPDE approximation. The comparisons we make are with respect to computational time, properties of estimators, posterior variability, and goodness fitness checking based on posterior predictive methods. Our objective is to provide practical guidance for users of the LGCP model. In addition to these comparisons, there are two novel aspects to the work presented here. First, we develop a mean field variational Bayes approximation that incorporates the Laplace method to deal with the non-conjugacy of the LGCP model. To the best of our knowledge this is the first time such an approximation has been developed for approximate Bayesian inference with the LGCP model. Second, we apply HMC for fully Bayesian inference and a novel aspect of our implementation is that HMC is used to update the decay (correlation parameter) associated with the latent Gaussian process. A result is that the sampling algorithm mixes very well and to our knowledge the development of the HMC algorithm in this context is the first of its kind.

The remainder of the paper proceeds as follows. Section 2 discusses various approaches for conducting Bayesian inference for the LGCP model. Section 3 presents a comparison of approaches through simulation studies, while Section 4 makes comparisons using two real point pattern datasets, the first arising from an ecological application and the second arising from a neuroimaging study. The paper concludes with a discussion in Section 5.

\section{Bayesian inference for log-Gaussian Cox processes}
\subsection{Model specification}
Consider an inhomogeneous Poisson process $\mathcal Z(s)$ with intensity function $\lambda(s)$, $s \in S \subseteq R^2$. Without loss of generality we shall assume that $S$ is the unit square. The density of a Poisson process does not exist with respect to Lebesgue measure, but the Radon-Nikodym derivative does exist with respect to a unit-rate Poisson process \citep{moller1998log}. We will call this derivative the density of the Poisson process. Given a set of $K$ points $\{s_k\} = \{s_1,...,s_K\} \subset S$, where both the number $K$ and the locations $s_k$
are random, the density is given by
\begin{equation}\label{intensity}
\pi[\{s_k \} \mid \lambda(s)]=\exp \left\{\int_S \left[ 1 -  \lambda(s)\right]ds \right\} \prod_{k=1}^K \lambda(s_k).
\end{equation}
where $\lambda(s)$ is the intensity function. If we further assume that the log of the intensity function arises from a Gaussian random field (GRF) $\mathcal Y(s)$ so that $\lambda(s)=\exp (\mathcal Y(s))$, then this hierarchical process is called a log-Gaussian Cox process (LGCP) \citep{moller1998log}. The LGCP, assumed to be stationary and isotropic, is uniquely determined by the mean function $\mu(s)$ and the covariance function $ \mbox{Cov}(s,s^\prime)=\sigma^2 r(||s-s^\prime||)$ of the Gaussian process $\mathcal Y(s)$, where $\sigma^2$ is the marginal variance and $r(||s-s^\prime||)$ denotes correlation as a function of the Euclidean distance $||s - s^\prime||$. Two commonly used correlation functions are the power exponential function \citep{moller2003statistical}
\begin{equation*}
 r_p(||s - s^\prime||)=\exp (-\rho  ||s - s^\prime||^\delta) 
\end{equation*}
where $\rho>0$ is the decay parameter, $\delta \in (0,2]$ is the power exponential term (which we will take as a known constant throughout this manuscript); and the Mat\'{e}rn correlation function \citep{matern1960spatial}
\begin{equation*}
r_m(||s - s^\prime||)={\left(\Gamma(\nu) 2^{\nu-1}\right)}^{-1} (||s - s^\prime|| /\phi)^{\nu}K_{\nu}(||s - s^\prime|| /\phi) 
\end{equation*}
where $\phi>0$ is the range parameter, $\nu>0$ is the shape parameter, and $K_{\nu}$ is the modified Bessel function of the second kind.

To fit the model in a tractable way a common approach is to divide the spatial domain into an $n \times n$ uniform grid of equally spaced cells \citep{moller1998log} and to make the simplifying assumption that the log-intensity is constant over each grid cell so that the log-intensity $\mathcal Y(s)$ within a given cell, say the $i^{th}$ cell, is constant and characterized by its value at the corresponding centroid, $c_i$, of cell $i$, $i \in \{1...n^2\}$. The unique log-intensity values then comprise a vector $\mathbf Y=(\mathcal Y(c_1),\mathcal Y(c_2),...,\mathcal Y(c_{n^2}))^{\mathrm T}$.  To simplify notation we let $Y_i = \mathcal Y(c_i)$ and $y_i$ is a realized value of $Y_i$. From the defining property of a GRF, $\mathbf Y$ follows a multivariate normal distribution $\mathbf Y \sim \mbox N(\mu \mathbf 1_{n^2},\sigma^2 \mathbf C)$, where $\mathbf C$ is the $n^2 \times n^2$ correlation matrix with elements $r(||c_i-c_j||)$. Let $\theta$ be the set of parameters determining the mean and covariance of the GRF  (e.g. $\theta=(\mu,\sigma^2,\rho)$ for the power correlation and  $\theta=(\mu,\sigma^2,\phi)$ for the Mat\'{e}rn), and let $A$ denote the area of each cell in the uniform grid. Under this discretization, the log density (see  Equation \eqref{intensity}) is
\begin{equation*}
\log \pi(\left\{s_k \right\} \mid \theta,\,\mathbf y) = \text{constant}+ \sum_{i} [ y_i n_i -A \exp( y_i)] 
\end{equation*}
where $n_{i}$ is the number of points in $\{s_k\}$ occurring in the $i^{th}$ grid cell. The log posterior can then be expressed as
\begin{eqnarray}
\log \pi(\theta,\, \mathbf y \mid \{s_k\}) &=& \text{constant}+ \sum_{i}[y_{i}n_i-A\exp(y_i)] \nonumber \\
 & &-0.5(\mathbf{y}-\mu \mathbf{1}_{n^2})^{\mathrm T}\sigma^{-2} \mathbf C^{-1}(\mathbf{y}-\mu \mathbf{1}_{n^2}) \nonumber \\
 && -0.5n^2\log(\sigma^2)-0.5\log(|\mathbf C |)+\log \pi(\theta)  \label{logpost}
\end{eqnarray}
where $\pi(\theta)$ is the prior density of the parameter vector $\theta$.
The computational problem for Bayesian inference is the calculation of $\pi(\theta,\, \mathbf y \mid \{s_k\})$ and its associated marginals or properties of these distributions. This computation is nontrivial because the calculation of the normalizing constant is nontrivial, particularly when the dimension of the parameter space is high. We now discuss three approaches for approximating the posterior distribution and/or its marginals.

\subsection{Hamiltonian Monte Carlo}

Hamiltonian Monte Carlo (HMC) has its origins with the work of \cite{alder1959studies} and \cite{duane1987hybrid} and was first introduced into the statistical literature by \cite{neal1995bayesian}. It is a Metropolis-Hastings algorithm that can be used for sampling a high-dimensional target distribution more efficiently than an algorithm based on random-walk proposals, especially when the parameters are highly correlated. The algorithm uses the notion of the (separable) Hamiltonian $H(\mathbf q, \mathbf p)$ from physics that is defined as the sum of potential energy $U(\mathbf q)$ and kinetic energy $K(\mathbf p)$, where $\mathbf q$ and $\mathbf p$ are random vectors that refer to position and momentum. The connection to Bayesian computation lies with relating $U(\mathbf q)$ to the posterior distribution and hence $\mathbf q$ to the model parameters, and with introducing auxiliary Gaussian random variables to represent momentum $\mathbf p$, a vector having the same length as $\mathbf q$. The evolution of this system is then described by the Hamilton equations from statistical mechanics:
\begin{eqnarray}\label{hamilton}
\frac{dq_i}{dt} &=& \frac{\partial H}{\partial p_i} \nonumber \\
\frac{dp_i}{dt} &=& -\frac{\partial H}{\partial  q_i}
\end{eqnarray}
which, if an analytic solution exists, produces a draw from the posterior distribution. In practice this system is solved using numerical integration techniques \citep{neal2011mcmc} and the resulting approximate solution is accepted or rejected using a Metropolis-Hastings step.

To carry out the computations required for the LGCP model we use a combination of re-parametrization of the random field and numerical techniques based on the 2D Fast Fourier transform (FFT) as in \cite{moller1998log}. Note that although \cite{girolami2011riemann} has applied RM-HMC to LGCP model, their computation is slow due to inverting the fisher information matrix and a Cholesky factorization of the correlation matrix. While here we can greatly speedup matrix multiplications involving the correlation matrix $\mathbf C$ (we note here that we use the power exponential correlation function) by using FFT. A second reason to use the re-parametrization is that we avoid inversion on the correlation matrix at each iteration, which in our simulations and data analyses below is a $4096\times 4096$ matrix.  Although this size of a matrix can be inverted on a computer, it is computationally expensive.  Furthermore, this FFT trick can handle much larger matrices that would be too large to invert on most computers.  To use this trick we require the matrix to be block-circulant as there is a direct relationship between the eigenvalue-eigenvector decomposition of a block-circulant matrix and the 2D discrete Fourier transform. However, the correlation matrix has a block-Toeplitz structure. A block-Toeplitz matrix can always be extended to a block-circulant matrix \citep{wood1994}. To do so, we extend the original $n\times n$ grid to an $m\times m$ grid and wrap it on a torus, where $m=2^g$ and $g$ is an integer such that $m \geq 2(n-1)$. The metric of this toroidal space is then defined by the minimum distance between two points. It is easy to show that the new correlation matrix $\mathbf E$ (of which $\mathbf C$ is a submatrix), whose elements are based on the metric defined on the torus, is a block-circulant matrix \citep{moller1998log}. In extending the space we must also expand the vector of latent variables $\mathbf Y$ in a corresponding manner, and we refer to this new vector as $\mathbf Y^{ext}$ (of which $\mathbf Y$ is a subvector). Also, we set the number of points in cell $i$, $m_i$, still to be $n_i$ if $i$ is on the original grid and equal to $0$ otherwise. 

The block-circulant extended correlation matrix can be decomposed as $\mathbf E=\mathbf F \Lambda \mathbf F^H$,  where $\mathbf F$ is the matrix of eigenvectors, $\Lambda$ is the diagonal matrix containing the corresponding eigenvalues of $\mathbf E$, and $H$ denotes the complex conjugate transpose. Given a random vector $\mathbf v$ of length $m^{2}$ the product $\mathbf E \mathbf v$ can be obtained by calculating $ \mathbf F^H \mathbf v$, $\Lambda \mathbf F^H \mathbf v$ and then $\mathbf F \Lambda \mathbf F^H \mathbf v$ in order. Note that the first and last calculations amount to a discrete inverse Fourier transform and a discrete Fourier transform (DFT), respectively. The middle calculation is simply element-wise multiplication of $\Lambda$ and the vector $\mathbf F^Hv$ \citep{rue2005gaussian}. As a result, the complexity of the required matrix operations can be reduced to $O(m^2\log(m^2))$ using the FFT. 

After extension of the grid we re-parametrize the latent variables $\mathbf Y^{ext}$ as $\mathbf Y^{ext}=\mu \mathbf 1_{m^2} + \sigma \mathbf E^{\frac12} \boldsymbol\gamma$ where $1_{m^2}$ denotes the $m^2$-dimensional vector of ones. $\boldsymbol\gamma=(\gamma_1,...,\gamma_{m^2})^{\mathrm T}$ with $\gamma_i \stackrel{iid}{\sim} \mbox N(0,1),i=1,...,m^2$. The gradients, used in the HMC algorithm, for all the parameters are straightforward to derive except for $\rho$, we give the expression for $\rho$ here and refer the readers to Appendix A for more details.

\begin{equation}
\frac{\partial \log \pi(\rho \mid \cdot)}{\partial \rho}=-\frac{\sigma}{2} \Big [  \mathbf m - A \exp \big ( \mu \mathbf 1_{m^2} + \sigma \mathbf E^{\frac12} \boldsymbol\gamma \big )  \Big ]^T \mathbf E^{-\frac12} \mathbf E^* \boldsymbol\gamma,
\end{equation}
where $\pi( \mid \cdot)$ denotes the full conditional given the data and other parameters, $\mathbf m$, $\mathbf E^*$ are defined in the appendix. And as $\mathbf E^{\frac12}$, $\mathbf E^{-\frac12}$, $\mathbf E^*$ are all block-circulant matrices, FFT can be used.

With the stochastic representation as in Equation \ref{hamilton} the HMC algorithm is based on setting $\mathbf U(\mathbf q) =- \log \left[\pi(\{s_k\} \mid \theta,\,\boldsymbol\gamma)\pi(\theta,\,\boldsymbol\gamma)\right]$ where 
$\mathbf q = (\boldsymbol\gamma^{\mathrm T},\theta^{\mathrm T})^{\mathrm T}$, and the kinetic energy term is $K(\mathbf p)= \mathbf p^{\mathrm T} \mathbf M^{-1} \mathbf p /2$, where $\mathbf M$ is a symmetric, positive-definite 'mass matrix' and auxiliary  momentum variables $\mathbf p$ (a vector of length $m^2+3$). In our work we set $\mathbf M$ to be a diagonal matrix with distinct diagonal components $m_{\boldsymbol\gamma}$, $m_{\mu}$, $m_{\sigma^{-2}}$ and $m_\rho$ corresponding to $\boldsymbol\gamma$ and $\theta$.  

Each iteration of the HMC algorithm involves a block update of $ \boldsymbol\gamma$ and  $\theta$ based on the Hamiltonian Monte Carlo scheme. Each such update requires $L+1$ evaluations of the gradient vector $\nabla_{\theta,\,\boldsymbol\gamma} \log \pi(\theta,\boldsymbol\gamma \mid \{s_k\})$ for some $L \ge 1$. If $L=1$, the HMC algorithm reduces to MALA, which typically mixes faster than the random walk Metropolis-Hastings algorithm \citep{roberts2001optimal} but not as fast as the more general HMC algorithm. Letting $\boldsymbol\gamma^*, \theta^*$ be the current value in the Markov chain for $\boldsymbol\gamma, \theta$, the HMC update, based on a step size $\varepsilon>0$, proceeds as follows: 
\begin{algorithm}
\caption{HMC algorithm}
\begin{enumerate}
\item Simulate latent vector $\mathbf p^* \sim \mbox N_{m^2+3}(\mathbf 0, \mathbf M)$.
Set 
\begin{eqnarray*}
\left(\boldsymbol\gamma^{(0)}, \theta^{0)}\right)&=&\left( \boldsymbol\gamma^*,\theta^*\right) \\
 \mathbf p^{(0)}  &=&\mathbf p^* + \frac{\varepsilon}{2} \nabla_{\theta^{*},\,\boldsymbol\gamma^{*}}\left[ \log\left\{ \pi\left( \boldsymbol\gamma^* ,\theta^* \mid \{s_k\}\right)\pi\left(\theta^*,\,\boldsymbol\gamma^*\right)\right\}\right].
 \end{eqnarray*}

\item For l=1,...,L, 
\footnotesize \begin{eqnarray*}
\left( \boldsymbol\gamma^{(l)}, \theta^{(l)}\right)^{\mathrm T} &=& \left( \boldsymbol\gamma^{(l-1)}, \theta^{(l-1)}\right)^{\mathrm T} + \varepsilon\mathbf M^{-1} \mathbf p^{(l-1)} \\
\mathbf p^{(l)} &=& \mathbf p^{(l-1)} + \varepsilon_l\nabla_{\theta^{(l)},\,\boldsymbol\gamma^{(l)}}\left[ \log\left\{ \pi\left( \boldsymbol\gamma^{(l)} ,\theta^{(l)} \mid \{s_k\}\right)\pi\left(\theta^{(l)},\,\boldsymbol\gamma^{(l)}\right)\right\}\right] 
\end{eqnarray*}
\normalsize\normalsize
where $\varepsilon_l=\varepsilon$ for $l<L$ and $\varepsilon_L=\varepsilon/2$. 

\item Accept $\left(\gamma^{(L)},\theta^{(L)}\right)$ as the new state for $(\gamma,\theta)$ with probability
\[
\alpha=\min \left( 1 , \exp\left\{-H\left(\mathbf q^{(L)}, \mathbf p^{(L)}\right)+H\left(\mathbf q^*, \mathbf p^*\right)\right\} \right)
\]
else remain in the current state $\gamma^*,\theta^*$ with probability $1-\alpha$.\\
\end{enumerate}
Repeat steps 1--3 for a sufficiently long time.
\end{algorithm}
At each iteration, the number of steps in the numerical integration, $L$, is drawn from a Poisson distribution with mean $100$, while the step size, $\varepsilon$, is initially chosen to be $0.005$ and adjusted adaptively during burning so that the acceptance rate is approximately $0.65$ \citep{beskos2013optimal}. Trace plots of the parameters are examined and based on these we adjust the values of $\mathbf M$ to improve the mixing.

\subsection{Mean field Variational Bayes with Laplace Approximation}
As an alternative to HMC (or MCMC) sampling of the posterior distribution, a deterministic approximation can be employed. Mean field variational Bayes (MFVB) is one such approximation that has been applied successfully to a number of problems, including spatial models for high-dimensional problems requiring fast computations \citep{nathoo2014variational}. For the MFVB algorithm, we return to the parametrization of the model in Equation \eqref{logpost}.  Let $q(\mathbf y, \theta)$ be an arbitrary density function having the same support as the posterior density $\pi(\theta,\,\mathbf y \mid \{s_k\})$. Letting $\log \pi(\{s_k\})$ denote the marginal likelihood of the model, we can express its logarithm as
\begin{eqnarray*}
\log \pi(\{s_k\}) &=& \int q(\mathbf y, \theta) \log \left\{  \frac{\pi(\{s_k\},\, \mathbf y,\, \theta)}{q(\mathbf y, \theta)}    \right\} \\ &&+ \int q(\mathbf y,\theta)\log \left\{ \frac{q(\mathbf y, \theta)}{\pi(\theta,\, \mathbf y  \mid \{s_k\})} \right \} \\
&\ge& \int q(\mathbf y,\theta) \log \left\{ \frac{\pi(\{s_k\},\,\theta,\,\mathbf y)}{q(\mathbf y,\theta)}   \right\}  \equiv F(q) 
\end{eqnarray*}
such that the functional $F(q)$ is a lower bound for $\log \pi(\{s_k\})$ for any $q$. The approximation is obtained by restricting $q$ to a manageable class of density functions, and maximizing $F$ over that class. We develop the approximation under the assumption that the GRF has a power exponential correlation function, and for now, we will assume that $\rho$ is known, so that $C$ is assumed known in what follows. 

We assume that the approximating density $q$ can be factorized 
\begin{equation}
\label{MF}
q(\mathbf y,\theta)=\left[\prod_{i=1}^{n^2} q(y_i)\right]q(\mu)q(\sigma^2).
\end{equation}
Under this assumption, a coordinate ascent algorithm is applied to maximize $F$ which leads to a sequence of coordinate-wise updates taking the form
\begin{eqnarray*}
&& q(y_i) \propto \exp \{ \mbox E_{-q(y_i)} [\log \pi(\{s_k\} \mid \mathbf y)\pi(\mathbf y \mid \theta)] \}  \quad \quad i=1,...,n^2 \label{q(y)} \\
&& q(\mu) \propto \exp \{\mbox E_{-q(\mu)} [\log \pi(\mathbf y \mid \theta) \pi(\mu)]\} \\
&& q(\sigma^2) \propto  \exp \{\mbox E_{-q(\sigma^{2})} [\log \pi(\mathbf y \mid \theta) \pi(\sigma^2)]\} 
\end{eqnarray*}
where $E_{-q(x)}[\cdot]$ denotes the expectation taken with respect to the set of random variables $\{\mathbf{y}, \theta\} \backslash x$ under the variational approximation $q_{-x}$, and the updates steps are iterated to the convergence of $F$. We describe the derivation of the update steps below. In what follows, $E[\cdot]$ will denote the expectation of its argument under the variational approximation $q$.

If conditionally conjugate Gaussian and inverse-gamma priors are chosen for $\pi(\mu)$ and $\pi(\sigma^2)$ respectively, $\mu \sim N(\mu_{\mu},\sigma_{\mu}^{2})$, $\sigma^{2} \sim G^{-1}(\alpha,\beta)$ the distributions $q(\mu)$ and $q(\sigma^2)$ comprising the update steps will also be Gaussian and inverse-gamma. To derive the update step for $\mu$ we have
\begin{equation}
\label{q_mu}
\begin{split}
 \mbox{E}_{-q(\mu)}\log \pi(\mathbf y \mid \theta) \pi(\mu)  =c-\frac{1}{2} \left[(\mathbf{\mu_y}-\mu \mathbf{1})^T \mbox{E}(\sigma^{-2}) (C^{-1}) (\mathbf{\mu_y} - \mu \mathbf{1})\right]-\frac{1}{2}(\mu-\mu_{\mu})^2/\sigma_{\mu}^2 \\
 =c- \frac{1}{2}\left[(\mbox{E}(\sigma^{-2}) \mathbf{1}^{T} (C^{-1}) \mathbf{1} + 1/\sigma_{\mu}^2 ) \mu^2 -2  \mbox{E}(\sigma^{-2}) \mathbf{\mu_y}^T (C^{-1}) \mathbf{1} \mu + \mu_{\mu} / \sigma^2_{\mu} \right ]
\end{split}
\end{equation}
where $\mathbf{\mu_y} = E[\mathbf y]$ and $c$ denotes a constant not depending on $\mu$. As (\ref{q_mu}) is a quadratic function of $\mu$ it follows that $q(\mu)$ is the density of a normal distribution $N(\mu_{\mu}^*, \sigma_{\mu}^{*2})$ where after some algebra we have
$$\mu_{\mu}^*=(\mbox{E}(\sigma^{-2}) \mathbf{\mu_y}^T (C^{-1})\mathbf{1}+ \mu_{\mu} / \sigma_{\mu}^2)(\sigma_{\mu}^{*2})^{-1},\,\,\,
\sigma_{\mu}^{*2}=(\mbox{E}(\sigma^{-2}) \mathbf{1}^{T} (C^{-1}) \mathbf{1} + 1/\sigma_{\mu}^2)^{-1}.
$$ 
To derive the update step for $\sigma^{2}$ we have

\begin{equation}
\begin{split}
\mbox E_{-q(\sigma^2)} \log \pi(\mathbf y \mid \theta) \pi(\sigma^2) =c -\frac{n^2}{2} \log \sigma^2 -\frac{1}{2} \sigma^{-2} \mbox{E}_{\mu} [ (\mathbf{\mu_y} - \mu \mathbf{1})^T C^{-1} (\mathbf{\mu_y} -\mu \mathbf{1}) \nonumber \\
+ \mbox{Tr}(C^{-1} \Sigma_{\mathbf{y}})]-(\alpha+1) \log \sigma^2 -\beta / \sigma^2  
\end{split}
\end{equation}
where $\Sigma_{\mathbf{y}}$ is the covariance matrix of $\mathbf y$ under $q(\mathbf y)$ and $c$ denotes a constant not depending on $\sigma^{2}$. Simplifying this expression yields
\begin{align}
\begin{split}
\mbox E_{-q(\sigma^2)} \log \pi(\mathbf y \mid \theta) \pi(\sigma^2) = &c-(\alpha+1+\frac{n^2}{2}) \log \sigma^2 - (\beta+\frac{1}{2}[(\mathbf{\mu_y} - \mu_{\mu}^* \mathbf{1})^T (C^{-1}) (\mathbf{\mu_y} - \mu_{\mu}^* \mathbf{1}) \nonumber \\
&+ \sigma_{\mu}^{*2} \mathbf{1}^{T} (C^{-1}) \mathbf{1}+\mbox{Tr}(C^{-1} \Sigma_{\mathbf{y}})])/\sigma^2
\end{split}
\end{align}
and thus $q(\sigma^2)$ is the density of an Inverse-Gamma distribution $G^{-1}(\alpha+\frac{n^2}{2},\beta^*)$ where $\beta^* = \beta +\frac{1}{2}\left[(\mathbf{\mu_y}-\mu_{\mu}^* \mathbf{1})^{T} (C^{-1}) (\mathbf{\mu_y} -\mu_{\mu}^* \mathbf{1}) + \sigma_{\mu}^{*2} \mathbf{1}^{T}  (C^{-1}) \mathbf{1}+ \mbox{Tr}(C^{-1} \Sigma_{\mathbf{y}})\right]$.

As the Gaussian prior for $y_i$ is not conditionally conjugate for the LGCP model, the variational Bayes update for $q(y_i)$ is not a standard distribution and is therefore not easy to compute without some further approximation. We derive an update step for $q(y_i)$ by applying the Laplace method \citep{wang2013variational} within the variational Bayes update. We have
\begin{align}
%\begin{split}
\mbox{E}_{-q(y_i)} \log \pi(\{s_k\} \mid \mathbf y)\pi(\mathbf y \mid \theta)= c+ (y_{i}n_{i}-e^{y_{i}}A) \nonumber\,\,\,\,\,\,\,\,\,\,\,\,\,\,\,\,\,\,\,\,\,\,\,\,\,\,\,\,\,\,\,\,\,\,\,\,\,\,\,\,\,\,\,\,\,\,\,\,\,\,\,\,\,\,\,\,\,\,\,\,\,\,\,\,\,\,\,\,\,\,\,\,\,\,\,& \\ 
-\frac{1}{2} \left[(\mathbf{\tilde{y}}-\mbox{E}(\mu) \mathbf{1})^T \mbox{E}(\sigma^{-2})(C^{-1}) (\mathbf{\tilde{y}} -\mbox{E}(\mu) \mathbf{1})+\mbox{Var}(\mu)E(\sigma^{-2})\mbox{Tr}((C^{-1}) \mathbf{1} \mathbf{1}^{T})\right]\nonumber&\\
=c+ (y_{i}n_{i}-e^{y_{i}}A)-\frac{1}{2}\left[(\mathbf{\tilde{y}}-\mbox{E}(\mu) \mathbf{1})^{T} \mbox{E}(\sigma^{-2}) (C^{-1}) (\mathbf{\tilde{y}} -\mbox{E}(\mu) \mathbf{1})\right]&
%\end{split}
\end{align}
where $\mathbf{\tilde{y}}$ denotes $(\mbox{E}(y_1),...,\mbox{E}(y_{i-1}),y_i,\mbox{E}(y_{i+1}),...,\mbox{E}(y_{n^2}))^{T}$.
Taking the derivative with respect to $y_i$ yields,
\begin{eqnarray} \label{eq:1}
 f(y_{i})= \partial \mbox{E}_{-q(y_i)} \log \pi(\{s_k\} \mid \mathbf y)\pi(\mathbf y \mid \theta)/\partial y_{i} = n_{i}-A e^{y_i}-\mbox{E}(\sigma^{-2})\left[(C^{-1})(\mathbf{\tilde{y}}-\mbox{E}(\mu) \mathbf{1})\right]_{i} \nonumber \\
=n_i-A e^{y_i} -\mbox{E}(\sigma^{-2}) \sum_{j} (C^{-1})_{ij}(\mathbf{\tilde{y}}-\mbox E (\mu) \mathbf{1} )_j  \nonumber \\  
=-A e^{y_i} - \mbox{E}(\sigma^{-2}) (C^{-1})_{ii} y_i +n_i+\mbox E(\sigma^{-2}) \mbox (C^{-1})_{ii} \mbox E(\mu) \mathbf - \mbox E(\sigma^{-2}) \sum_{j\neq i} \mbox (C^{-1})_{ij} (\mathbf{\tilde{y}}- \mbox E(\mu) \mathbf 1)_j
\end{eqnarray}
Given ($\ref{eq:1}$) we find $\hat y_i$ such that $f(\hat y_i)=0$. We use Newton's method to obtain a numerical solution where the starting value for Newton's method is obtained by omitting the linear term in equation ($\ref{eq:1}$) and solving the resulting simplified equation exactly. We then take the second derivative with respect to $y_i$,
$
H(y_{i}) = \partial f(y_{i})/\partial y_{i}=-A e^{y_i}-\mbox{E}(\sigma^{-2})(C^{-1})_{ii}
$
and given this and the solution $\hat y_i$, the Laplace method yields a normal distribution  $N(\hat{y}_i,-H(\hat{y}_i)^{-1})$ for $q(y_{i})$ which approximates the VB update. The variational-Laplace approximation to the posterior of the latent field $\mathbf{y}$ is then $\mathbf{y} \sim N(\mathbf{\mu_y},\Sigma_{\mathbf{y}})$ where $\mathbf{\mu_y}=(\hat{y}_1,...,\hat{y}_{n^2})^{T}$ and $\Sigma_{\mathbf{y}}$ is a diagonal matrix with the $i^{th}$ element being $-H(\hat{y}_i)^{-1}$.

Given the update steps derived above, the approximate posterior distribution under the variational-Laplace approximation takes the form \eqref{MF} where the component densities are standard distributions
\begin{eqnarray*}
\mathbf y &\sim& \mbox N(\mu_{\mathbf y}^{(q)}, \Sigma_{\mathbf y}^{(q)}) \\
\mu &\sim& \mbox N(\mu_{\mu}^{(q)},\sigma_{\mu}^{2(q)}) \\
\sigma^2 &\sim& G^{-1} (\alpha^{(q)},\beta^{(q)}).
\end{eqnarray*}
The parameters determining these distributions are called the 'variational parameters'. These parameters are obtained through the sequence of update steps derived above which are used to determine equations expressing each variational parameter in terms of the remaining variational parameters; beginning with initial values for these parameters the equations are iterated to convergence of $F$. 

With respect to the parameter $\rho$ of the power exponential correlation function, we have found that its inclusion as an unknown parameter into the variational approach leads to convergence problems in a number of trial examples. To deal with this problem, we estimate this parameter prior to running the VB algorithm using the method of minimum contrast \citep{diggle1984monte}, i.e., to use non-linear least squares estimation to fit a non-parametric estimated covariance function. The mean field VB algorithm incorporating the Laplace method for the LGCP model is presented in detail in Algorithm 2.

\begin{algorithm}[htbp]
\caption{Mean Field VB Algorithm with Laplace Method}
\begin{enumerate}
\item Initialize the priors $\mu_{\mu},\sigma^2_{\mu},\alpha,\beta$. 
\item Initialize $\mu_{\mathbf y}^{(q)},\,\Sigma_{\mathbf y}^{(q)},\mu_{\mu}^{(q)},\sigma_{\mu}^{2(q)},\beta^{(q)}$ and $\mbox E(\sigma^{-2})=(\alpha+n^2/2)/(\beta^{(q)})$
\item Obtain $\rho$ using the minimum contrast method, compute $\mathbf C^{-1}$ where $c_{kl}=\exp(-\rho ||c_k - c_l ||^\delta)$. 
 
\item For $i=1,...,n^2$, compute $\mu_{y_i}^{(q)}$ such that 
 \begin{eqnarray*}
 \lefteqn{-A \exp(\mu_{y_i}^{(q)}) - \mbox{E}(\sigma^{-2}) \left({\mathbf C}^{-1}\right)_{ii} \mu_{y_i}^{(q)} +n_i} \\
 &&+\mu_{\mu}^{(q)} \mbox E(\sigma^{-2}) \left({\mathbf C}^{-1}\right)_{ii}  \mathbf - \mbox E(\sigma^{-2}) \sum_{j\neq i} \left({\mathbf C}^{-1}\right)_{ij}[\mu_{\mathbf y}^{(q)}- \mu_{\mu}^{(q)} \mathbf 1]_j =0
 \end{eqnarray*}
 where $[\cdot]_j$ denotes the $j^{th}$ element of a vector.\\
Compute $H(\mu_{y_i}^{(q)})=-A \exp(\mu_{y_i}^{(q)})-\mbox{E}(\sigma^{-2}) \left({\mathbf C}^{-1}\right)_{ii}$. \\Obtain $\mu_{\mathbf y}^{(q)}$ and $\Sigma_{\mathbf y}^{(q)}$ where $\mu_{\mathbf y}^{(q)}=(\mu_{y_1}^{(q)},...,\mu_{y_{n^2}}^{(q)})^{\mathrm T}$ and $\Sigma_{\mathbf y}^{(q)}$ is a diagonal matrix with diagonal elements $-H(\mu_{y_i}^{(q)})^{-1}$.

\item Compute $\mu_{\mu}^{(q)}=(\mbox{E}(\sigma^{-2}) {\mu_{\mathbf y}^{(q)}}^{\mathrm T} \mathbf C^{-1} \mathbf{1}+\mu_{\mu} / \sigma_{\mu}^{2})(\mbox{E}(\sigma^{-2}) \mathbf{1}^{\mathrm T} \mathbf C^{-1} \mathbf{1} + 1/\sigma_{\mu}^{2})^{-1}$. 
Compute $\sigma_{\mu}^{2(q)}=(\mbox{E}(\sigma^{-2}) \mathbf{1}^{\mathrm T} \mathbf C^{-1} \mathbf{1} + 1/\sigma_{\mu}^2)^{-1}$. 
\item Compute $\beta^{(q)} = \beta +0.5[(\mu_{\mathbf y}^{(q)}-\mu_{\mu}^{(q)} \mathbf{1})^{\mathrm T} \mathbf C^{-1} (\mu_{\mathbf y}^{(q)} -\mu_{\mu}^{(q)} \mathbf{1}) + \sigma_{\mu}^{2(q)} \mathbf{1}^{\mathrm T} \mathbf C^{-1} \mathbf{1}+ \mbox{Tr}(\mathbf C^{-1} \Sigma_{\mathbf y}^{(q)})]$.  
Obtain $\mbox{E}(\sigma^{-2})=(\alpha+n^2/2)/(\beta^{(q)})$ 

\item Compute the lower bound 
\begin{eqnarray*}
F(q) &=& \sum_i (\mu_{y_i}^{(q)} n_i - A \exp(\mu_{y_i}^{(q)}-0.5H(\mu_{y_i}^{(q)} )^{-1}))-0.5 \mbox E \log (|\mathbf C| )  \\
&&-0.5 [(\mu_{\mu}^{(q)} - \mu_{\mu})^2 +\sigma_{\mu}^{2(q)}]/ \sigma_{\mu}^2 \\ && +\sum_i 0.5 \log \left(-H(\mu_{y_i}^{(q)})^{-1}\right) +0.5 \log \sigma_{\mu}^{2(q)}
\end{eqnarray*}
\end{enumerate}
Repeat 4--7 until the increase in $F(q)$ is negligible.
\end{algorithm}

\subsection{INLA}
The integrated nested Laplace approximation (INLA) is another approach for constructing a deterministic approximation to the posterior distribution that can be applied to the fairly broad class of latent Gaussian models. The details underlying the approach have been described in a number of recent papers including the seminal work of \cite{rue2009approximate}. We provide here only a brief overview of aspects that are relevant for use with the LGCP model. 

For spatial models, INLA makes extensive use of the Gaussian Markov random field (GMRF) which is a Gaussian distribution having a sparse precision matrix. Algorithms for fitting models incorporating a GMRF can be made efficient through the use of numerical methods for sparse matrices. In the case of the LGCP model the latent GRF which is a spatially continuous process is approximated by a GMRF on a discrete lattice so that these numerical methods can be applied. We consider this approximation for the case where the GRF is a Mat\'{e}rn field with $\nu$ known, so that $\theta=(\mu,\sigma^2,\phi)$ and the log intensity is characterized by $\mathbf Y$ as before. An approximation $\tilde \pi(\theta \mid \{s_k\})$ to the marginal posterior distribution $\pi(\theta \mid \{s_k\})$ is first obtained using the Laplace method. An approximation $\tilde \pi(y_i \mid \theta, \{s_k\})$ to the density of the full conditional distribution of each component of the latent field is then obtained using one of the three methods: a Gaussian approximation, the Laplace approximation, or a simplified Laplace approximation. The latter option is based on a series expansion of the Laplace approximation which has a lower computational cost. The marginal posterior distributions for the log intensity values of the LGCP model are then approximated through numerical integration over a discrete grid for $\theta$
\[
\tilde \pi (y_i \mid \{s_k\}) = \sum_k \tilde \pi(y_i \mid \theta_k, \{s_k\}) \tilde \pi (\theta_{k}, \{s_k\}) \Delta_k
\]
where $\{\Delta_k\}$ are a set of area weights associated with the grid.

Recently, \cite{lindgren2011explicit} develop an approximation to certain GRFs with Mat\'{e}rn correlation functions by specifying stochastic partial differential equations (SPDEs) that have certain Mat\'{e}rn processes as their solution. This SPDE representation provides an explicit link to GMRFs through a basis function representation of the solution where the corresponding weights comprise a GMRF with dependencies determined by a triangular mesh covering the spatial domain. This approximation can also be embedded within INLA and is implemented within the R-INLA package (obtained at www.r-inla.org) which allows for different mesh sizes. Increasing the size of the mesh will increase the accuracy of this approximation but will also increase the required time for computation.

We note, and express, here that the INLA package only allows for a Mat\'{e}rn correlation structure, whereas for the HMC and VB algorithms we use the power exponential family for computational purposes (e.g.~it allows for easy calculation of the gradients).

\section{Simulation Studies}

Here the methods described in the previous section are compared using simulation studies. The discretized spatial domain is taken to be a $64 \times 64$ grid on the unit square, so that the simulated data are based on $4096$ spatial locations. Each study is based on 1000 datasets simulated under the discretized LGCP model where we compare a total of six approaches: HMC incorporating FFT methods on the extended grid, VB incorporating the Laplace method, INLA with a simplified Laplace approximation (INLA I), INLA with a full Laplace approximation (INLA II), INLA with the SPDE model based on a mesh size of 436 (INLA III), and INLA with the SPDE model based on a mesh size of 4075 (INLA IV). The first mesh size was chosen purposely small and the second mesh size was chosen to be approximately the  same as the number of cells in the discretized grid.  In what follows, INLA I and INLA II are also referred to as INLA with the lattice method, INLA III and INLA IV are also referred to as INLA with SPDE.

Our HMC and VB algorithms were derived under the assumption of a power exponential correlation for the GRF; whereas, we use the implementation of INLA in the standard R-INLA package that assumes a Mat\'{e}rn correlation. When simulating data we assume the GRF has a Mat\'{e}rn correlation and we apply all six approaches to the resulting data. Thus, INLA is based on a correctly specified covariance function and therefore has an advantage over HMC and VB which have the correlation function misspecified. We will also assume that the parameter $\nu$ (in the Mat\'{e}rn model) is fixed and known, so that using INLA we only estimate the decay parameter $\phi$ in the Mat\'{e}rn model. As we cannot directly compare $\rho$ and $\phi$ we are not able to compare directly the properties of the corresponding estimators. As an alternative we make comparisons with respect to the distance at which the correlation function drops to $0.5$, denoted as $d_{0.5}$ and defined by the equations
\begin{equation}
r_p(d_{0.5})=r_m(d_{0.5})=0.5
\end{equation}
where $r_p(\cdot)$ and $r_m(\cdot)$ denote the power exponential and Mat\'{e}rn correlation functions respectively.

\subsection{Simulation One}
We first simulate datasets from the LGCP model where the Mat\'{e}rn field has $\mu=5$, $\sigma^2=3.5$, $\phi=0.02$, and $\nu=1$. Based on these parameters we simulate the GRF once and, based on this realization of the latent field, we simulate 1000 independent replicates of the data. Along with estimation of the log intensity values we also estimate $\mu, \sigma^2$, $d_{0.5}$ and $\mbox E(N)$, where $\mbox E(N)$ is the expected total number of points of the process within the spatial domain. The estimators are evaluated with respect to bias, variance and MSE. For the HMC and VB algorithms we match the shape of the power correlation function with that of the Mat\'{e}rn correlation function based on nonlinear least squares to estimate, and fix,  the value for $\delta$ in the power exponential model, and we obtain a value of  $\delta = 1.312$. 

In terms of priors, HMC assumes a flat prior $\pi(\mu) \propto 1$, $\sigma^2 \propto \mbox I(0,\infty)$ and $\rho \propto \mbox I(0,\infty)$. With VB we are constrained to use conditionally conjugate priors and set $\mu \sim \mbox N(0,625)$, $\sigma^2 \sim G^{-1}(1,1)$, while $\rho$ is estimated by the method of minimum contrast and assumed known in the VB algorithm. For INLA with the lattice method, we assign diffuse priors $\sigma^{-1} \sim G(0.001,0.001)$ and $d_{I} \sim G(0.001,0.001)$ where $d_{I}=\sqrt{8 \nu }\phi$,  and for INLA with SPDE we use the default joint-normal prior. Additional discussion of the priors and related numerical issues associated with INLA SPDE are mentioned in Section 5. INLA III has a mesh size of $436$ (based on a length $0.1$ for the inner mesh and $0.5$ for outer mesh) and INLA IV has a mesh size of $4075$ (based on a length $0.03$ for the inner mesh and $0.5$ for outer mesh). We acknowledge that the use of different priors for the HMC, VB, and INLA methods is not ideal as the differences we observe in the simulation results may, to some extent, be driven by differences in the priors. However, as the priors are taken to be fairly diffuse in all cases we do not expect the differences in the priors to play a significant role. As a practical matter the form of the prior may sometimes be driven by the choice of computational algorithm used for Bayesian computation. Indeed, convergence issues may also impact the prior used in some circumstances. While not ideal from a theoretical perspective these issues are unavoidable from a practical standpoint. For example, the use of VB typically calls for conditionally conjugate priors while for INLA we are constrained to use certain forms for the priors that are built into the R-INLA package. As the different computational algorithms are often implemented with different priors we feel that these comparisons offer useful practical guidance for users despite these differences. 

\begin{figure}[!ht]
\begin{center}
\includegraphics[width=\linewidth]{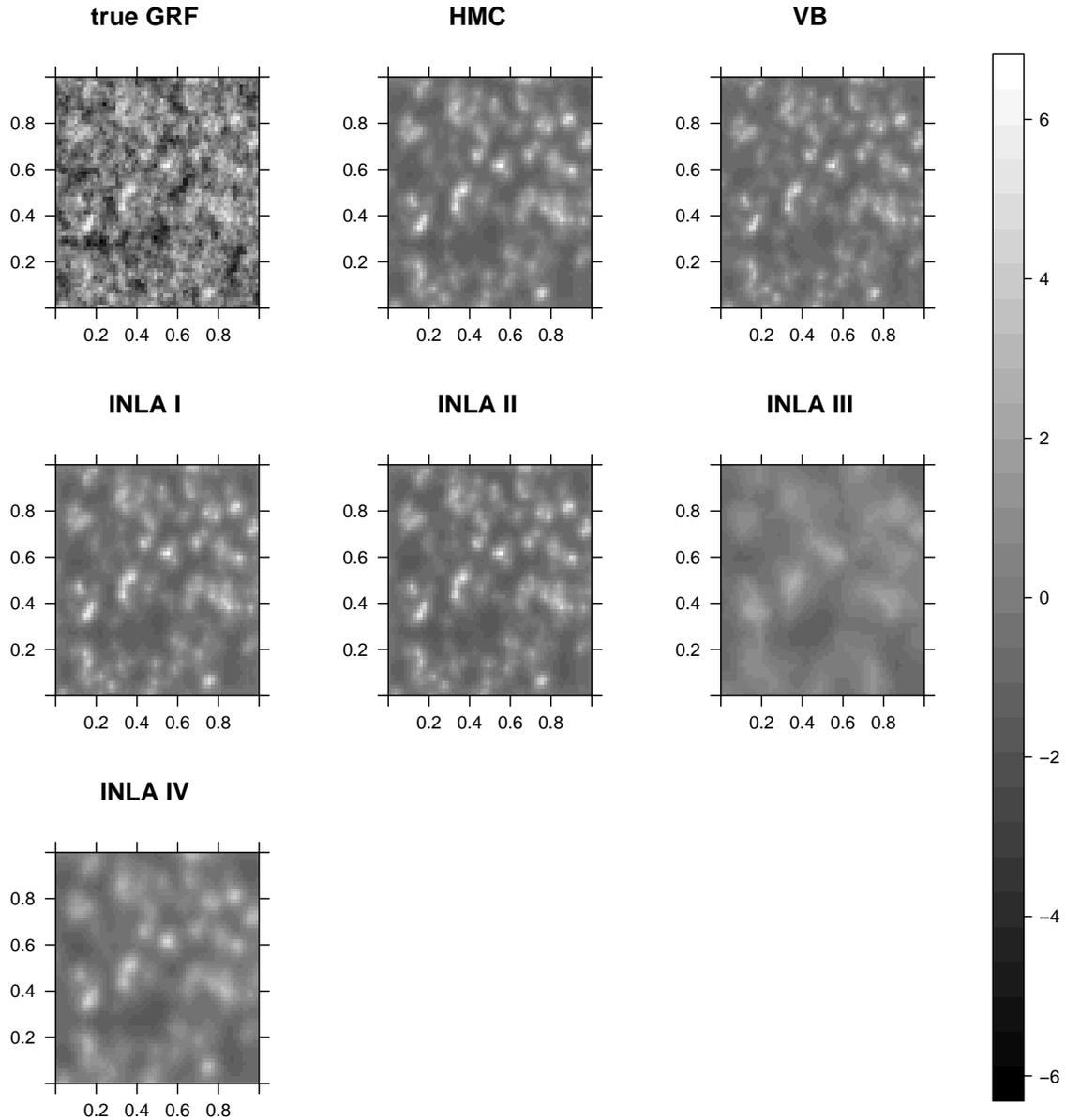}
\caption{Average marginal posterior mean of the log-intensity over 1000 samples from the first simulation study. Upper left panel is the true GRF.
} \label{fig1}
\end{center}
\end{figure}

Figure \ref{fig1} shows the true value of the discretized latent field $\mathbf Y$ in comparison to the average marginal posterior mean obtained from each of the methods under consideration. The average marginal posterior mean is the average of the estimates obtained from each of the 1000 simulation replicates. In this case we see that HMC, VB, INLA I and INLA II all produce similar average reconstructions of $\mathbf y$; whereas, the results from INLA III and INLA IV appear more over-smoothed, with the degree of over-smoothing decreasing as the mesh size increases.

\begin{figure}[!ht]
\begin{center}
\includegraphics[width=\linewidth]{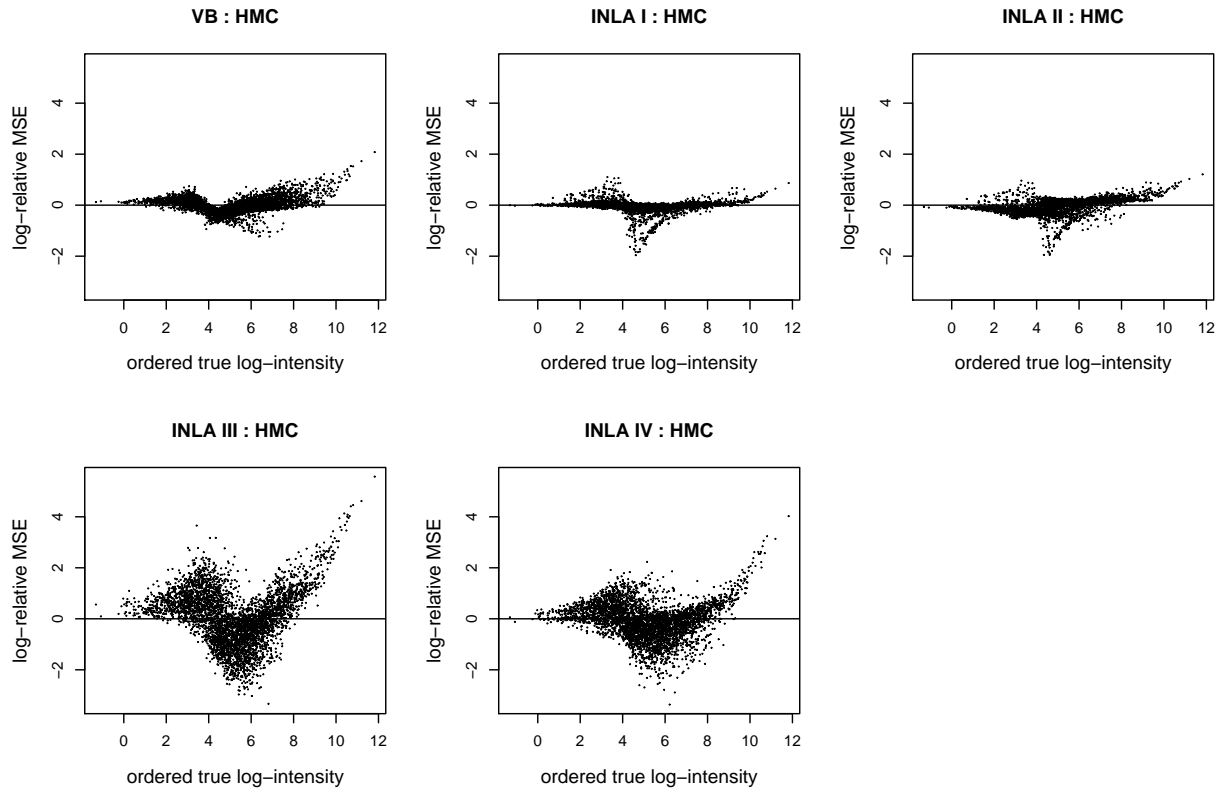}
\caption{Log-relative MSE of estimated latent GRF from VB and INLA I--IV to that from HMC for the first simulation study. Each point represents the log-relative MSE of the discretized GRF. Points located above, on and below the horizontal line denotes bigger, equal and smaller MSE than that from HMC.
} \label{fig2}
\end{center}
\end{figure}

Taking HMC as the baseline, a plot of the log-relative mean squared error (MSE) associated with the posterior mean estimator of $\mathbf Y$ for VB and INLA is shown in Figure \ref{fig2}. Points above (below) the black line indicate larger (smaller) MSE relative to that obtained from HMC. Both VB and INLA I-IV tend to produce estimators that have a higher MSE than the corresponding estimator obtained from HMC when the value of the log-intensity approaches either tail of the distribution of log-intensity values. Conversely, the methods appear to outperform HMC for values of the log-intensity around the median of this distribution. These differences appear to be more variable for INLA III and INLA IV compared with VB, INLA I and INLA II.

Table \ref{propertysim1} displays the bias, variance, and MSE for the posterior mean estimators of $\mu, \sigma^{-2}$, $d_{0.5}$ and $\mbox E(N)$. In this and all subsequent tables we display, for HMC, the actual values of bias, variance and MSE, whereas for VB and INLA the values are relative to those obtained from HMC. For estimation of these parameters we see that VB and INLA generally have larger bias and MSE compared with HMC. Both INLA I and INLA II outperform VB slightly, while INLA III and INLA IV lag behind the alternatives rather significantly.  By construction, INLA I and INLA II will provide identical estimates for $\sigma^{-2}$ and $d_{0.5}$ and this is reflected in the table. For the estimation of $\mbox E(N)$, HMC outperforms VB with respect to bias, variance, and MSE, while these measures are not reported for INLA as we only obtain marginal posterior distributions of the log-intensity values in this case and thus cannot estimate $\mbox{E}(N)$.

\begin{table}[!ht]
\begin{center}\footnotesize
\begin{tabular}{llr|rrrrr}
\hline \hline
\multicolumn{1}{c}{Parm}	&	\multicolumn{1}{c}{Measure}	& \multicolumn{1}{c}{}		&	\multicolumn{5}{c}{Relative Measure} \\
	&		&	\multicolumn{1}{c}{HMC}	&	\multicolumn{1}{c}{VB}	&	\multicolumn{1}{c}{INLA I}	&	\multicolumn{1}{c}{INLA II}	&	\multicolumn{1}{c}{INLA III}	&	\multicolumn{1}{c}{INLA IV}	\\ \hline
$\mu=5$	&	Bias	&	0.046	&	12.681	&	5.469	&	8.471	&	25.805	&	15.435	\\
	&	Var	&	0.010	&	0.476	&	0.903	&	0.663	&	8.453	&	0.632	\\
	&	MSE	&	0.012	&	29.441	&	6.143	&	13.507	&	127.227	&	43.558	\\ \hline
$\sigma^{-2}=0.286$	&	Bias	&	-0.010	&	6.984	&	-3.727	&	-3.727	&	-72.906	&	-14.545	\\
	&	Var	&	3.20E-04	&	0.454	&	1.338	&	1.338	&	36.531	&	4.810	\\
	&	MSE	&	4.08E-04	&	11.436	&	4.191	&	4.191	&	1236.295	&	51.801	\\ \hline
$d_{0.5}=0.025$	&	Bias	&	3.30E-05	&	-122.467	&	122.245	&	122.245	&	1704.511	&	1115.598	\\
	&	Var	&	2.20E-06	&	0.465	&	1.244	&	1.244	&	1199.323	&	4.574	\\
	&	MSE	&	2.20E-06	&	7.895	&	8.588	&	8.588	&	2626.682	&	616.258	\\ \hline
$\mbox E(N)=910.29$	&	Bias	&	-1.337	&	-252.070	&	-	&	-	&	-	&	-	\\
	&	Var	&	918.282	&	1.420	&	-	&	-	&	-	&	-	\\
	&	MSE	&	920.069	&	124.820	&	-	&	-	&	-	&	-	\\ \hline \hline
\end{tabular}
\end{center}
\caption{Summary of the statistical properties for the hyper-parameters from the first simulation study. The values shown in table from VB and INLA I-IV are relative to that from HMC.}
\label{propertysim1}
\end{table}

While Table \ref{propertysim1} displays the variance of the posterior mean estimators, we display in Table \ref{varsim1} the average (over simulation replicates) marginal posterior variance. The associated large sample theory guarantees that the posterior variance as obtained from HMC is simulation consistent. As such the posterior variance for VB and INLA are again listed relative to that obtained from HMC in order to determine the extent to which these approaches under-estimate or over-estimate posterior variability. For VB we see that the marginal posterior variance is under-estimated which is inline with expectations from the literature \citep{nathoo2013comparing,nathoo2014variational}. INLA I and II provide measures of variability that are closer to that of HMC, while INLA III and IV tend to over-estimate the marginal posterior variance, with this over-estimation being substantial when the smaller mesh size is used. With respect to average computational time, HMC requires 679 seconds based on 1500 total iterations with the first 500 thrown away as burn-in; VB requires 453 seconds to run to convergence (about 300 iterations); INLA I runs for 46 seconds, INLA II requires 196 seconds, INLA III requires 10 seconds, and INLA IV requires 154 seconds. The algorithms are run on an iMac with a 3.2GHz Intel Core i5 processor and 16GB memory.

\begin{table}[!ht]
\begin{center}\footnotesize
\begin{tabular}{lr|rrrrr}
\hline \hline
\multicolumn{2}{c}{Average Marginal Var}	 & \multicolumn{5}{c}{Relative Ave. Marg. Var}	\\
\multicolumn{1}{c}{Parm} &\multicolumn{1}{c}{HMC} 	&	\multicolumn{1}{c}{VB}	&	\multicolumn{1}{c}{INLA I}	&	\multicolumn{1}{c}{INLA II}	&	\multicolumn{1}{c}{INLA III}	&	\multicolumn{1}{c}{INLA IV}	\\ \hline
$\mu$	&	0.028	&	0.511	&	0.868	&	0.333	&	85.232	&	2.287	\\ \hline
$\sigma^{-2}$	&	0.001	&	0.029	&	1.576	&	1.576	&	6.780	&	17.061	\\ \hline
$d_{0.5}$	&	4.80E-06	&	0.022	&	1.667	&	1.667	&	87549.788	&	7.161	\\ \hline \hline
\end{tabular}
\end{center}
\caption{Marginal variance estimates of the parameters for the first simulation study. VB and INLA I-IV are relative to HMC}
\label{varsim1}
\end{table}

\subsection{Simulation Two}

\begin{figure}[!ht]
\begin{center}
\includegraphics[width=\linewidth]{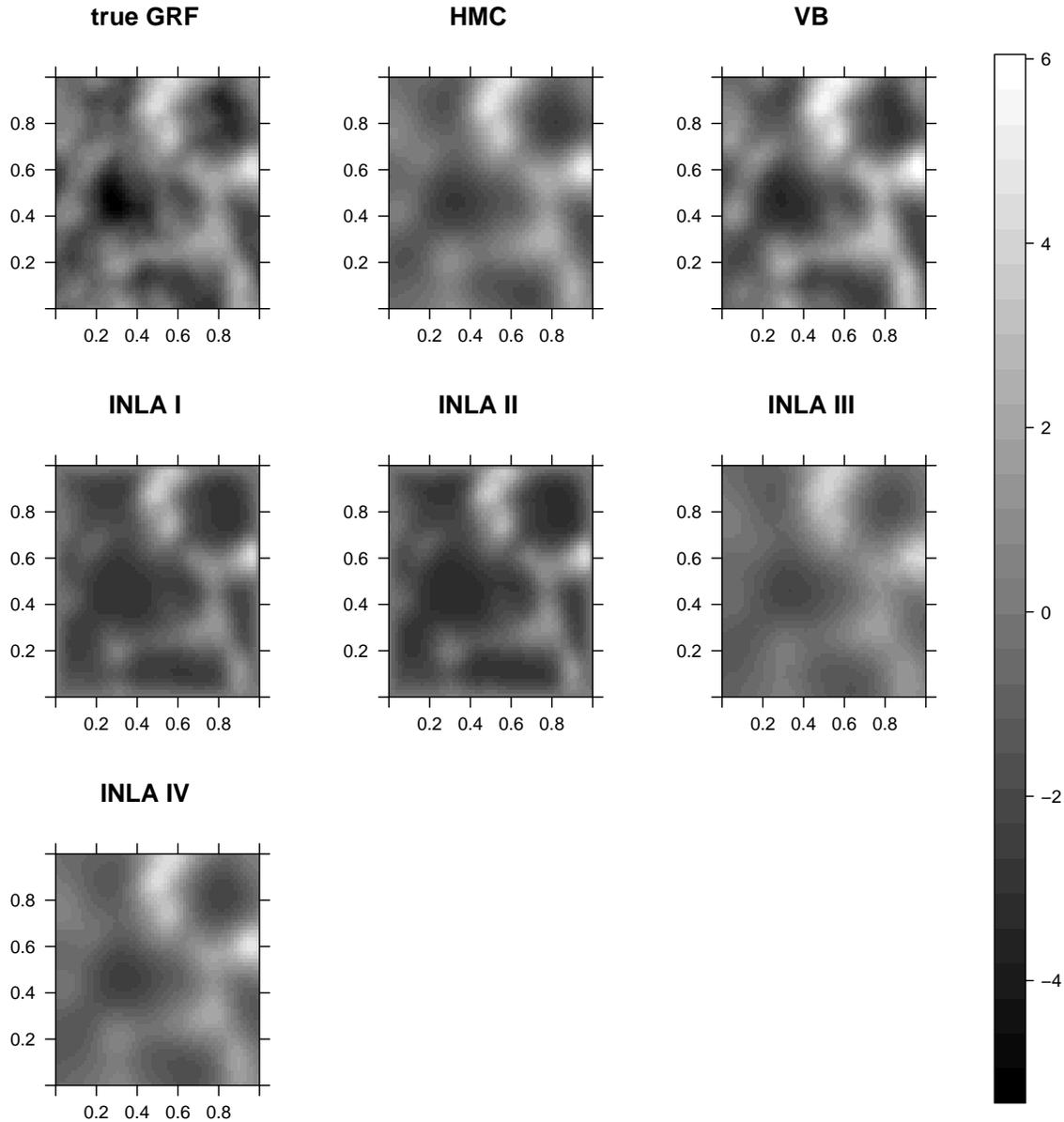}
\caption{Average marginal posterior mean of the log-intensity over 1000 samples from the second simulation study. Upper left panel is the true GRF.
} \label{fig3}
\end{center}
\end{figure}

\begin{figure}[!ht]
\begin{center}
\includegraphics[width=\linewidth]{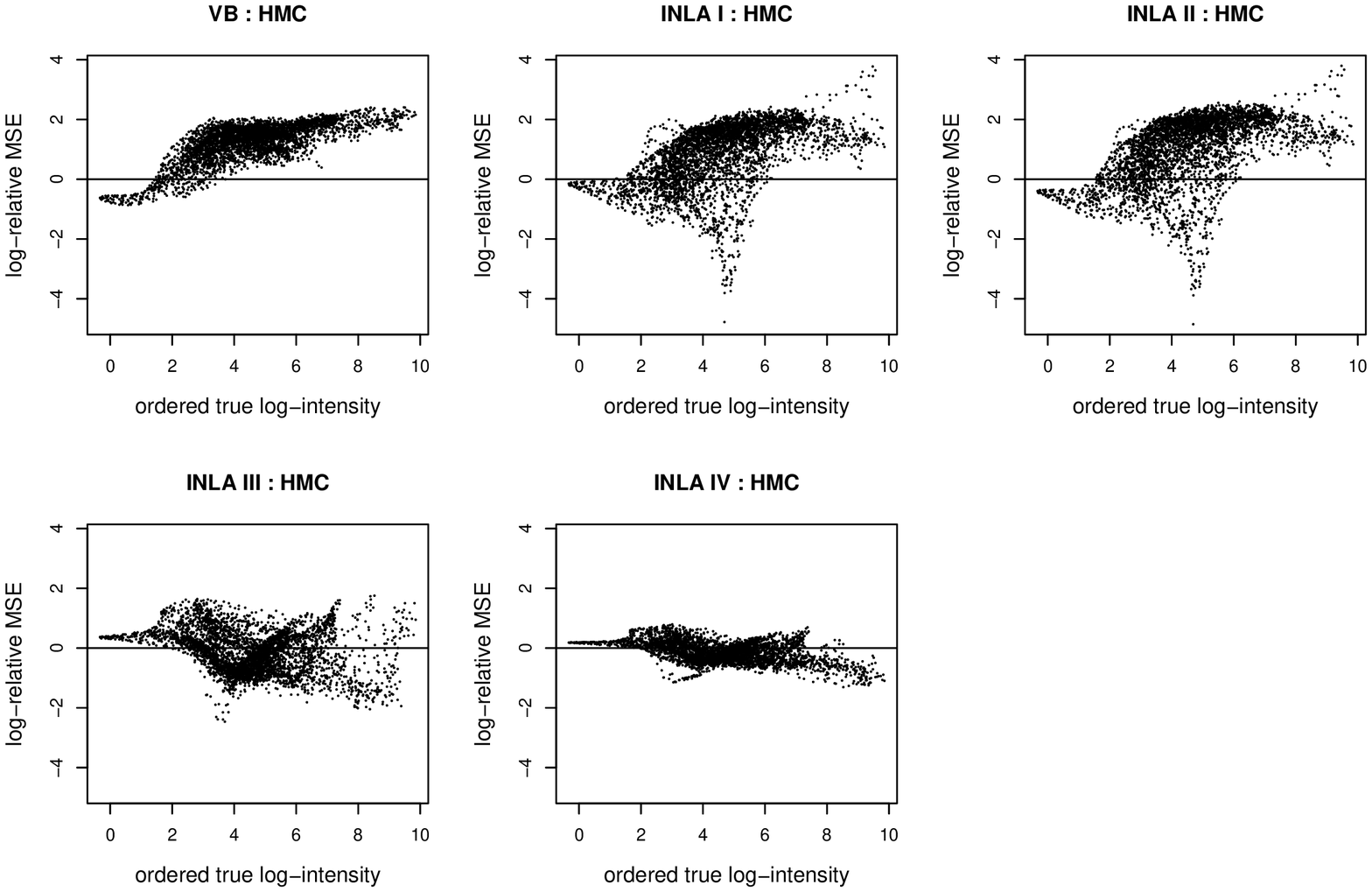}
\caption{Log-relative MSE of estimated latent GRF from VB and INLA I--IV to that from HMC for the second simulation study. Each point represents the log-relative MSE of the discretized GRF. Points located above, on and below the horizontal line denotes bigger, equal and smaller MSE than that from HMC.
} \label{fig4}
\end{center}
\end{figure}

In order to make comparisons in a setting where there is a slower decay for the spatial correlation and smoother realizations of the random field our second study is based on setting  $\phi=0.05$ and $\nu=3$ with all other settings remaining unchanged. Figure \ref{fig3} presents the average posterior mean log-intensity over 1000 simulation replicates. Comparing the images the overall best reconstruction seems to arise from both VB and HMC, followed by INLA I, INLA II, and INLA IV all of which capture the general features of the true latent field, while INLA III seems subject to over-smoothing as before. Figure \ref{fig4} displays the log-relative MSE of the five methods in comparison with HMC as in Figure \ref{fig2}. Interestingly, INLA IV seems to have the best performance in terms of MSE here, which taken together with Figure \ref{fig3} suggests that the estimators from INLA IV may have lower variance while still achieving adequate bias.

With respect to the hyper-parameters, Table \ref{propertysim2} displays the bias, variance, and MSE of the posterior mean estimators for $\mu,\sigma^{-2}$, $d_{0.5}$ and $\mbox E(N)$. INLA III and INLA IV have the smallest MSE for $\mu$, but have large MSE for $d_{0.5}$. As before HMC attains the lowest MSE for estimation of $\sigma^{-2}$ and $d_{0.5}$. In terms of $\mbox E(N)$, HMC outperforms VB with respect to bias, variance and MSE. Table \ref{varsim2} shows the average marginal posterior variance for the hyper-parameters. INLA I and II generally under-estimate the marginal posterior variance. VB shows significant under-estimation of the marginal variance for $\sigma^{-2}$ and $d_{0.5}$, and over-estimation of the posterior variance for $\mu$. As with the previous study, INLA III and INLA IV  over-estimate the marginal posterior variance for all three hyper-parameters. In terms of average timing, HMC requires 521 seconds for a total of 2000 iterations with the first 1000 iterations discarded as burn-in; VB requires 467 seconds and typically required a greater number of iterations to converge (about 1000) compared to simulation one; INLA I takes 134 seconds, INLA II takes 357 seconds, INLA III takes 11 seconds, INLA IV takes 227 seconds.

\begin{table}[!ht]
\begin{center}\footnotesize
\begin{tabular}{llr|rrrrr}
\hline \hline
\multicolumn{1}{c}{Parm}	&	\multicolumn{1}{c}{Measure}	&	   \multicolumn{1}{c}{}    &	\multicolumn{5}{c}{Relative Measure}\\
	&		&	\multicolumn{1}{c}{HMC}	&	\multicolumn{1}{c}{VB}	&	\multicolumn{1}{c}{INLA I}	&	\multicolumn{1}{c}{INLA II}	&	\multicolumn{1}{c}{INLA III}	&	\multicolumn{1}{c}{INLA IV}	\\ \hline
$\mu=5$	&	Bias	&	-0.458	&	2.752	&	-1.513	&	-1.488	&	0.047	&	0.458	\\
	&	Var	&	0.029	&	1.169	&	4.682	&	4.617	&	0.932	&	1.232	\\
	&	MSE	&	0.239	&	6.797	&	2.579	&	2.507	&	0.115	&	0.334	\\ \hline
$\sigma^{-2}=0.286$	&	Bias	&	-0.017	&	15.316	&	3.355	&	3.355	&	2.038	&	4.477	\\
	&	Var	&	1.20E-03	&	0.003	&	3.75	&	3.75	&	1.131	&	0.849	\\
	&	MSE	&	0.001	&	48.492	&	5.301	&	5.301	&	1.755	&	4.817	\\ \hline
$d_{0.5}=0.13$	&	Bias	&	-1.00E-02	&	4.364	&	6.893	&	6.893	&	-43.224	&	-31.924	\\
	&	Var	&	7.90E-05	&	0.191	&	1.236	&	1.236	&	24.439	&	17.627	\\
	&	MSE	&	1.80E-04	&	10.868	&	27.431	&	27.431	&	1068.207	&	584.571	\\ \hline
$\mbox E(N)=494.12$	&	Bias	&	0.077	&	1201.419	&	-	&	-	&	-	&	-	\\
	&	Var	&	496.455	&	1.199	&	-	&	-	&	-	&	-	\\
	&	MSE	&	496.461	&	18.527	&	-	&	-	&	-	&	-	\\ \hline	\hline						
\end{tabular}
\end{center}
\caption{Summary of the statistical properties for the hyper-parameters from the second simulation study. The values shown in table from VB and INLA I-IV are relative to that from HMC.}
\label{propertysim2}
\end{table}

\begin{table}[!ht]
\begin{center}\footnotesize
\begin{tabular}{lr|rrrrr}
\hline \hline
\multicolumn{2}{c}{Average Marginal Var}	 & \multicolumn{5}{c}{Relative Ave. Marg. Var}	\\
\multicolumn{1}{c}{Parm} &\multicolumn{1}{c}{HMC} 	&	\multicolumn{1}{c}{VB}	&	\multicolumn{1}{c}{INLA I}	&	\multicolumn{1}{c}{INLA II}	&	\multicolumn{1}{c}{INLA III}	&	\multicolumn{1}{c}{INLA IV}	\\ \hline$\mu$	&	0.255	&	6.084	&	0.301	&	0.299	&	5.952	&	4.531	\\ \hline
$\sigma^{-2}$	&	0.006	&	3.36E-05	&	0.650	&	0.650	&	23.322	&	20.661	\\ \hline
$d_{0.5}$	&	3.10E-04	&	0.001	&	0.164	&	0.164	&	12.023	&	5.935	\\ \hline \hline
\end{tabular}
\end{center}
\caption{Marginal variance estimates of the parameters from the second simulation study. VB and INLA I-IV are relative to HMC}
\label{varsim2}
\end{table}

\section{Application}

We next compare the computational algorithms through an application to two datasets where the LGCP model is applied in both cases. In addition to comparing the methods with respect to posterior summaries of the parameters of interest, we also make comparisons with respect to goodness-of-fit checking using the posterior predictive distribution \citep{gelman1996posterior} which has been applied for checking hierarchical spatial models in a number of applications including disease ecology and neuroimaging  \citep[e.~g.]{nathoo2010joint,kang2011meta}). The posterior predictive checks are based on the $L$ function \citep{illian2009hierarchical} where we simulate, based on the model, posterior predictive replicates of the discrepancy measure $\Delta(r)=L(r,\{s_k\}^{obs},\mathbf y, \theta)-L(r,\mathbf \{s_k\}^{rep},\mathbf y,\theta)$ where $\mathbf \{s_k\}^{obs}$ denotes the observed data, $\mathbf y$ and $\theta$ are drawn from the posterior distribution, and $\mathbf \{s_k\}^{rep}$ denotes replicate data that is drawn from the posterior predictive distribution. For a given distance range $r$, if the value $\Delta(r) = 0$ is observed to lie as an extreme value in either tail of the posterior predictive distribution we may question the fit of the model as characterized by the $L$ function at that distance. As INLA does not provide the joint posterior distribution we are unable to simulate predictive realizations and thus we make the predictive comparisons comparisons only between HMC and VB.

All priors are the same as in the simulation studies unless otherwise indicated. We also compare VB and the INLA methods with HMC as HMC is simulation consistent.

\subsection{Bramble Canes data} 
\begin{figure}[!ht]
\begin{center}
\begin{subfigure}{0.41\textwidth}
\includegraphics[width=\linewidth]{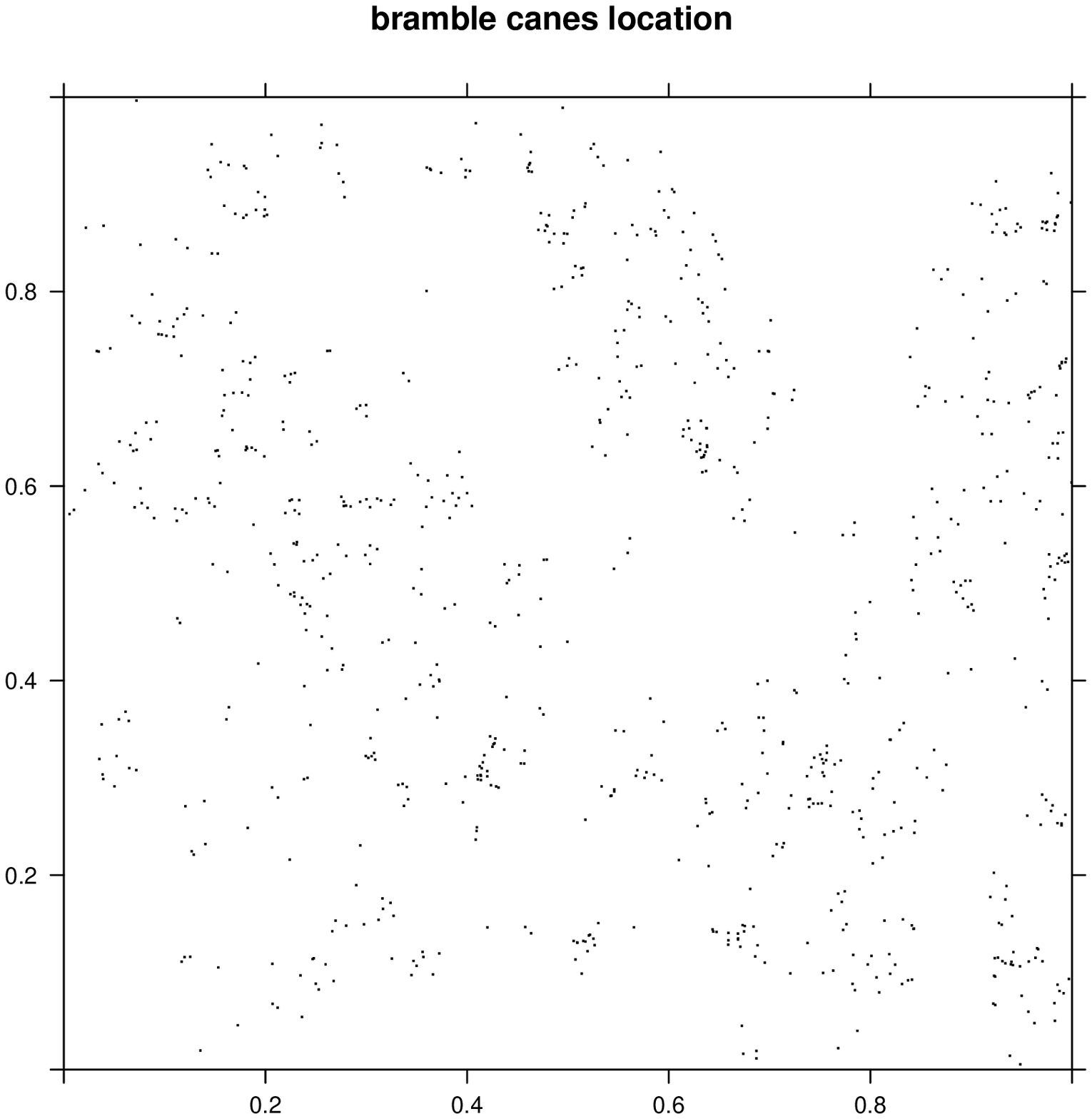}
\caption{} \label{fig5a}
\end{subfigure}
\begin{subfigure}{0.41\textwidth}
\includegraphics[width=\linewidth]{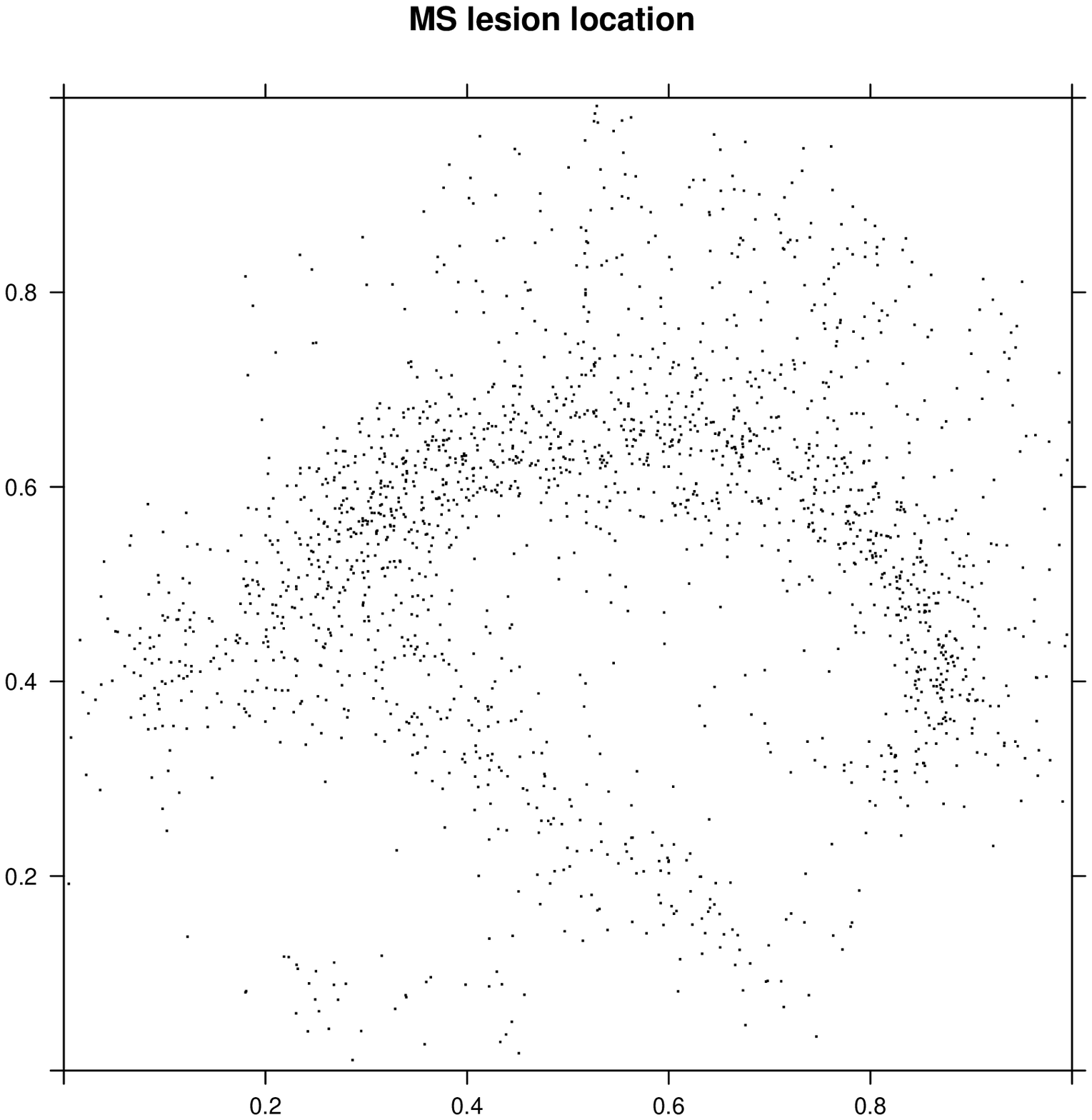}
\caption{} \label{fig5b}
\end{subfigure}
\caption{(a) shows the Bramble canes locations; (b) shows the MS lesion locations. Both are represented by black dots, rescaled to the unit square} 
\label{fig5}
\end{center}
\end{figure}

The data record the $(x,y)$ locations of $823$ bramble canes in a field of $9\,\mbox{m}^2$, rescaled to a unit square. The data are depicted in Figure \ref{fig5}(a) and were recorded and analyzed by Hutchings (1979) and further analyzed by Diggle \citep{diggle1983statistical}. 

\begin{figure}[!ht]
\begin{center}
\includegraphics[width=\linewidth]{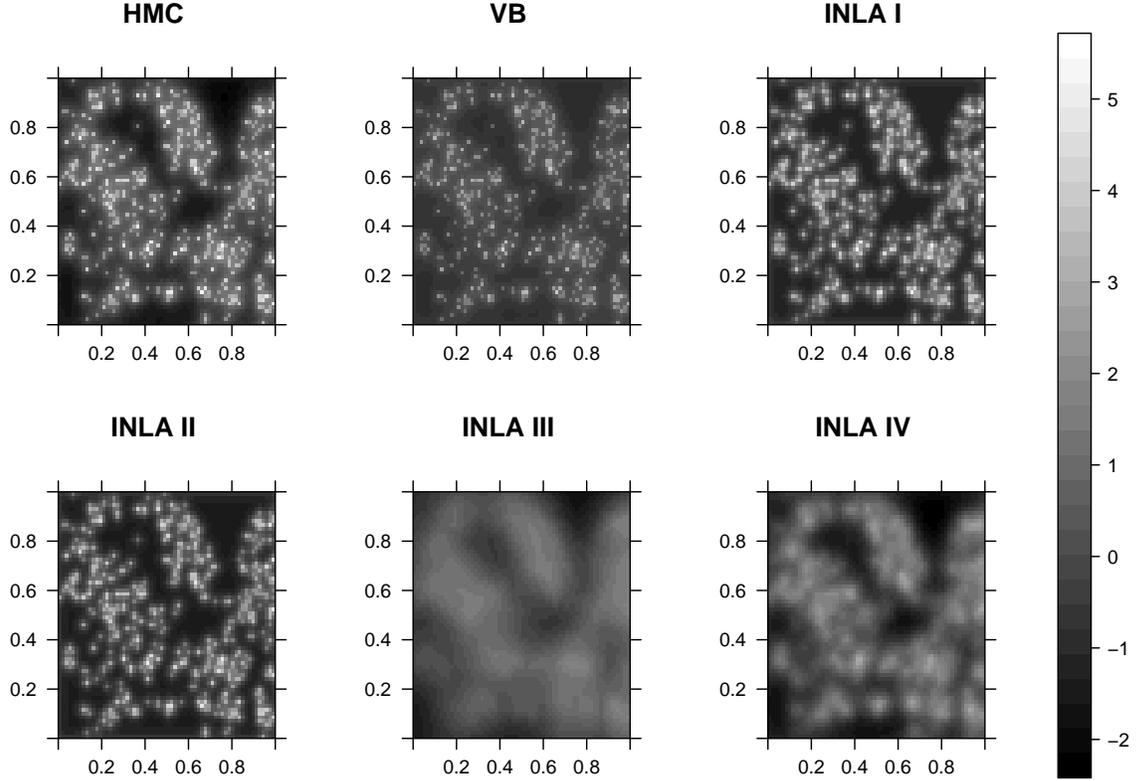}
\caption{Posterior mean of the latent GRF for the bramble canes data set, estimated from HMC, VB and INLA I--IV.
} \label{fig6}
\end{center}
\end{figure}

To determine the value of $\delta$ in the power-exponential correlation function and the value of $\nu$ in the Mat\'{e}rn correlation function, we use the method of minimum contrast which estimates the values as  $\hat{\delta}=0.51$, $\hat{\nu}=0.02$.  As the R-INLA package only offers three possible values $\nu=1,2,3$, we select $\nu=1$ as it is the closest of the three choices to the estimate obtained from the minimum contrast method. Figure \ref{fig6} depicts the posterior mean of the log-intensity as obtained by each of the six methods. HMC, INLA I and INLA II result in estimated images that appear fairly consistent; results obtained from VB are also consistent with HMC but less so than INLA I and II, while both INLA III and INLA IV appear to over-smooth the estimated latent field relative to the other methods. 

\begin{figure}[!ht]
\begin{center}
\includegraphics[width=\linewidth]{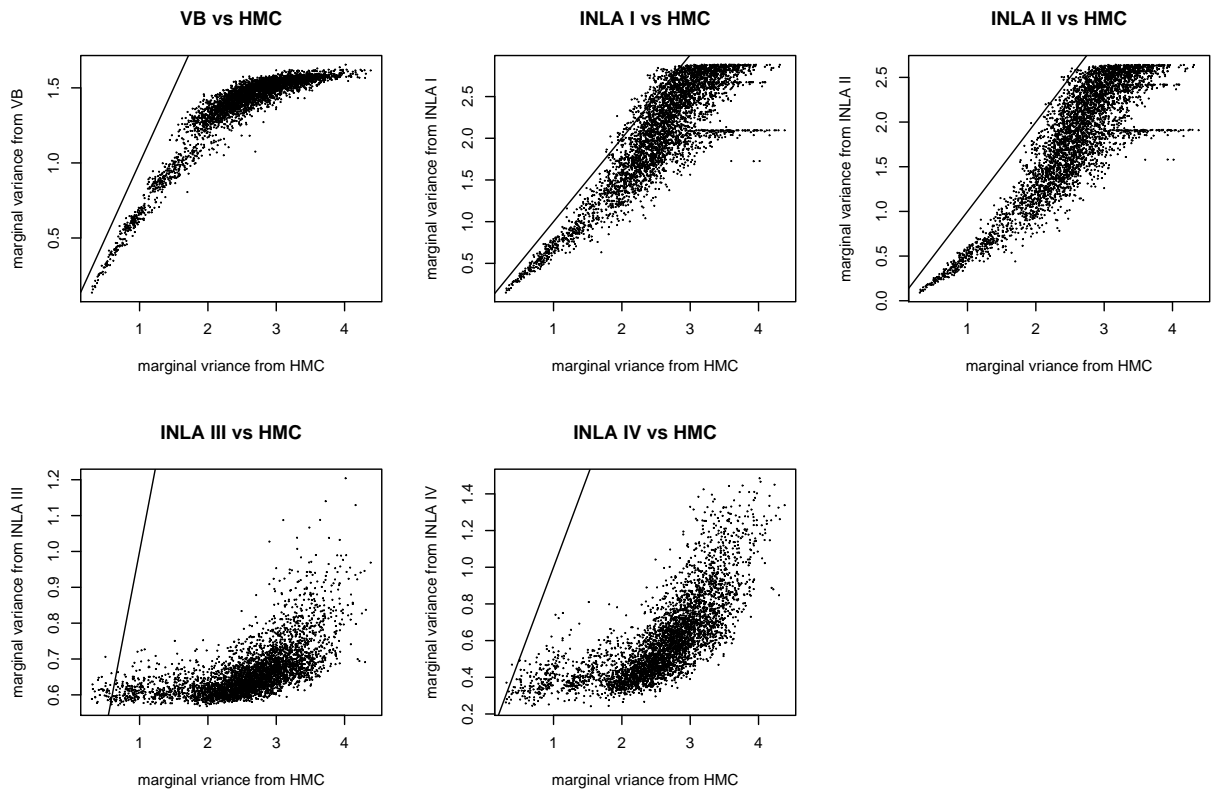}
\caption{Scatter plot of the marginal posterior variance of the latent GRF from VB  and INLA I-IV compared with those from HMC. Bramble canes data set.
} \label{fig7}
\end{center}
\end{figure}

Posterior summaries of the hyper-parameters $\mu,\sigma^{-2}$ and $d_{0.5}$ are presented in Table \ref{varreal1}. Here we see that VB is under-estimating the posterior variance for all hyper-parameters while the point estimates for $\mu,\sigma^{-2}$ are larger than those obtained from HMC, but within somewhat reasonable bounds. The point estimates obtained from INLA with the lattice method are closer to those obtained from HMC than those obtained from INLA with SPDE. In Figure \ref{fig7} we compare the marginal posterior variance of each element of the latent field as obtained from all of the methods to the posterior variance obtained from HMC. In this case all of the methods under-estimate the posterior variance and this under-estimation is most severe for INLA with SPDE.

\begin{table}[!ht]
\begin{center}\footnotesize
\begin{tabular}{llr|rrrrr}
\hline \hline
\multicolumn{1}{c}{Parm}	&	\multicolumn{1}{c}{Measure}	&	\multicolumn{1}{c}{}          &\multicolumn{5}{c}{Relative Measure}		\\
	&		&	\multicolumn{1}{c}{HMC}	&	\multicolumn{1}{c}{VB}	&	\multicolumn{1}{c}{INLA I}	&	\multicolumn{1}{c}{INLA II}	&	\multicolumn{1}{c}{INLA III	}&	\multicolumn{1}{c}{INLA IV}	\\ \hline
$\mu$	&	Post. Mean	&	5.019	&	1.223	&	0.996	&	1.078	&	1.197	&	1.161	\\
	&	Post. Var	&	0.016	&	0.52	&	1.503	&	0.535	&	32.469	&	4.744	\\ \hline
$\sigma^{-2}$	&	Post. Mean	&	0.272	&	1.743	&	1.135	&	1.135	&	2.461	&	1.998	\\
	&	Post. Var	&	0.001	&	0.194	&	1.581	&	1.581	&	114.182	&	26.765	\\ \hline
$d_{0.5}$	&	Post. Mean	&	0.025	&	0.165	&	0.768	&	0.768	&	9.215	&	2.894	\\
	&	Post. Var	&	8.00E-05	&	9.98E-06	&	0.025	&	0.025	&	46.779	&	0.945	\\ \hline \hline
\end{tabular}
\end{center}
\caption{Summary of parameter estimation for the bramble canes data set, VB and INLA I-IV are relative to HMC.}
\label{varreal1}
\end{table}

\begin{figure}[!ht]
\begin{center}
\begin{subfigure}{0.41\textwidth}
\includegraphics[width=\linewidth]{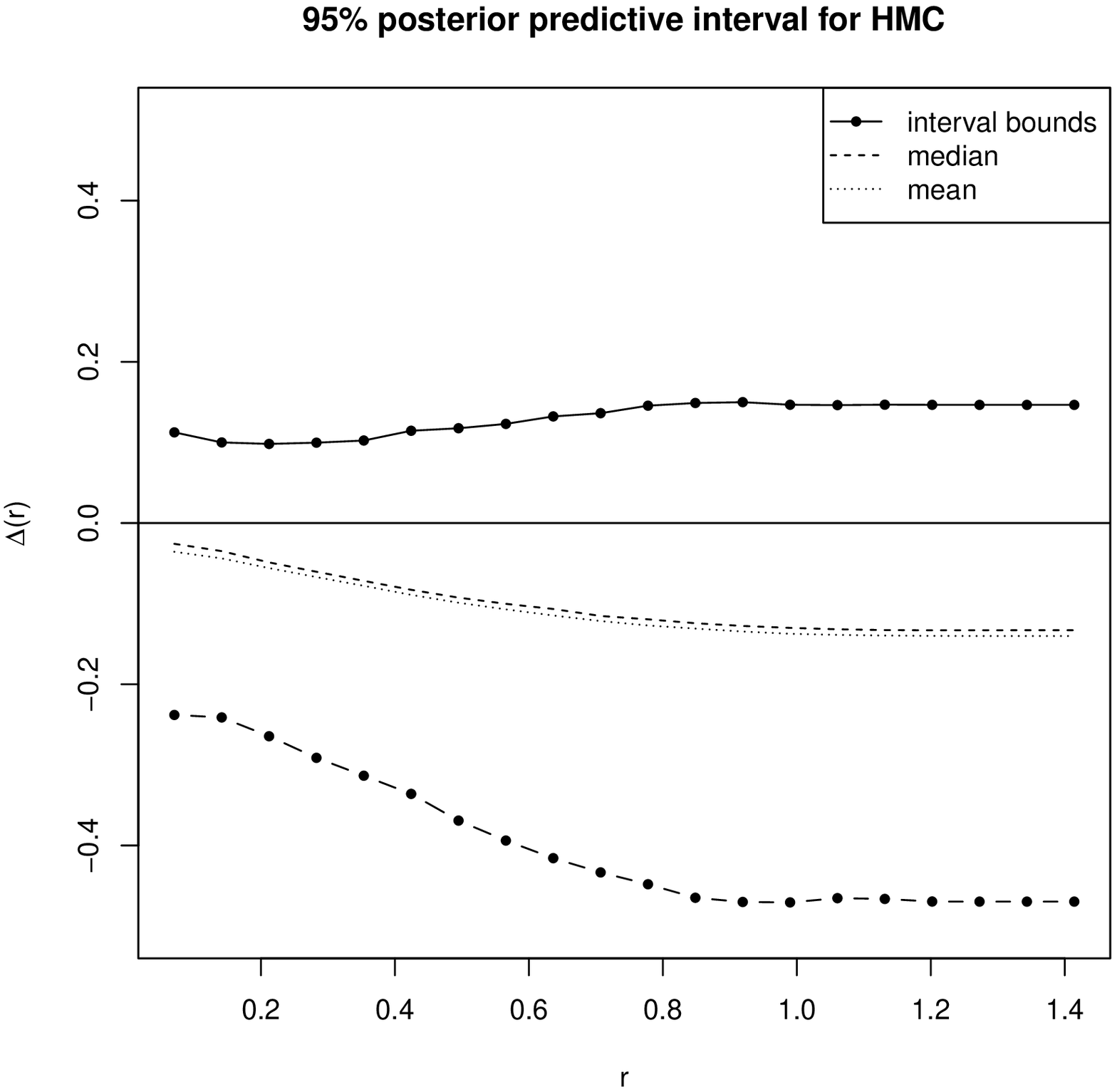}
\caption{} \label{fig8a}
\end{subfigure}
\begin{subfigure}{0.41\textwidth}
\includegraphics[width=\linewidth]{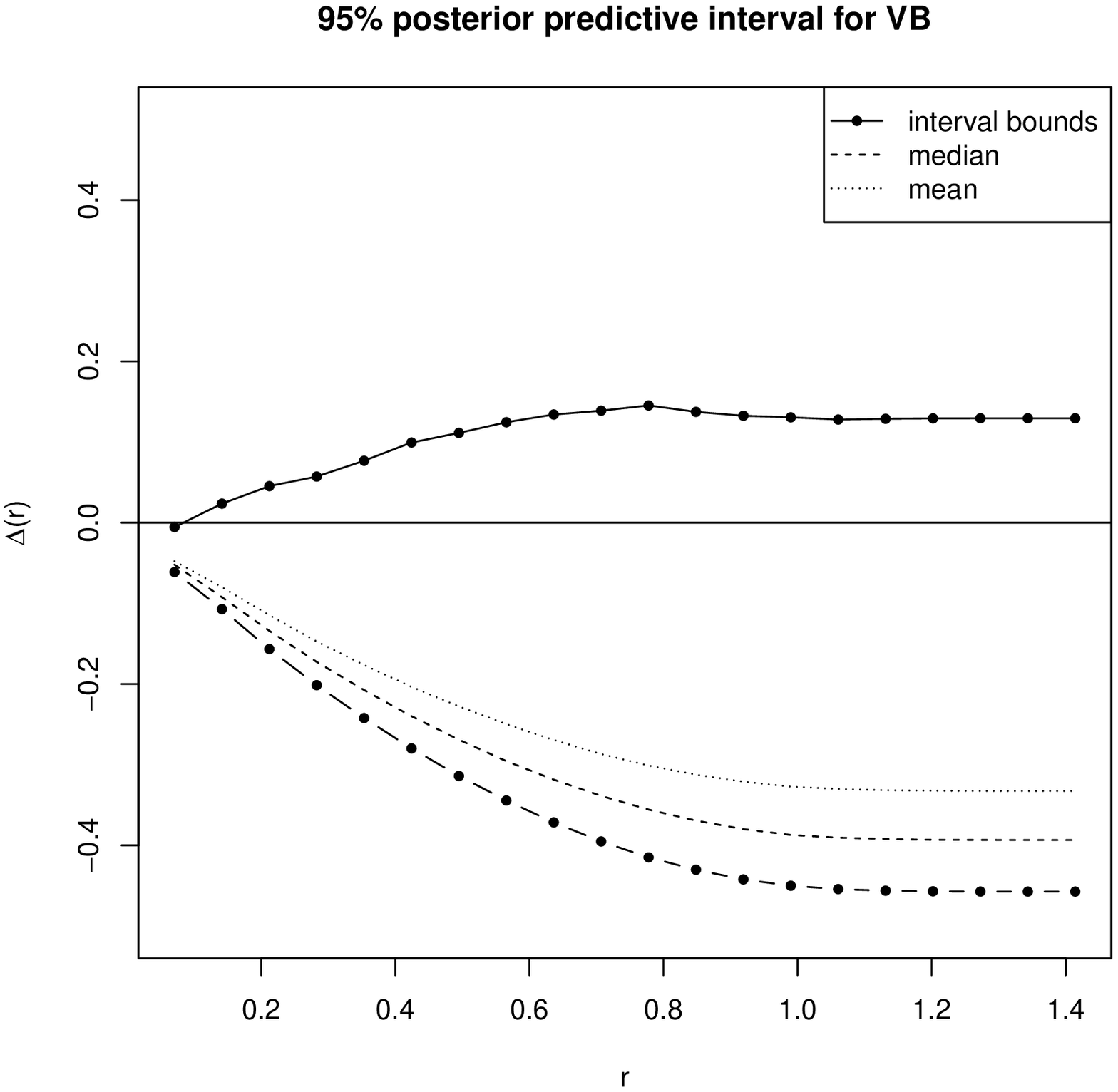}
\caption{} \label{fig8b}
\end{subfigure}
\caption{95\% posterior predictive interval for HMC (a) and VB (b) for the bramble canes data set. The bounds are denoted by solid lines while the mean and median are denoted by dashed lines. These are obtained at 20 distinct distances.} 
\label{fig8}
\end{center}
\end{figure}

Figure \ref{fig8} compares the 95\% posterior predictive intervals for $\Delta(r)$ as obtained from both HMC and VB. Comparing the figures indicates that the posterior predictive variability is under-estimated for VB, primarily at the lower distance ranges $r$. The implication of this for data analysis is that posterior predictive checks for VB under similar settings would be conservative. In terms of the data, the posterior predictive check as obtained from HMC does not reveal a lack-of-fit with respect to the chosen discrepancy measure. With respect to timing, HMC requires 597s for 1500 iterations and 500 burn-in iterations; VB actually requires more time than HMC in this example and takes 1012s requiring 137 iterations to convergence. INLA I require 55s, INLA II 172s, INLA III 11s, and INLA IV takes 227s.

\subsection{Multiple Sclerosis MRI Data}
\begin{figure}[!ht]
\begin{center}
\includegraphics[width=\linewidth]{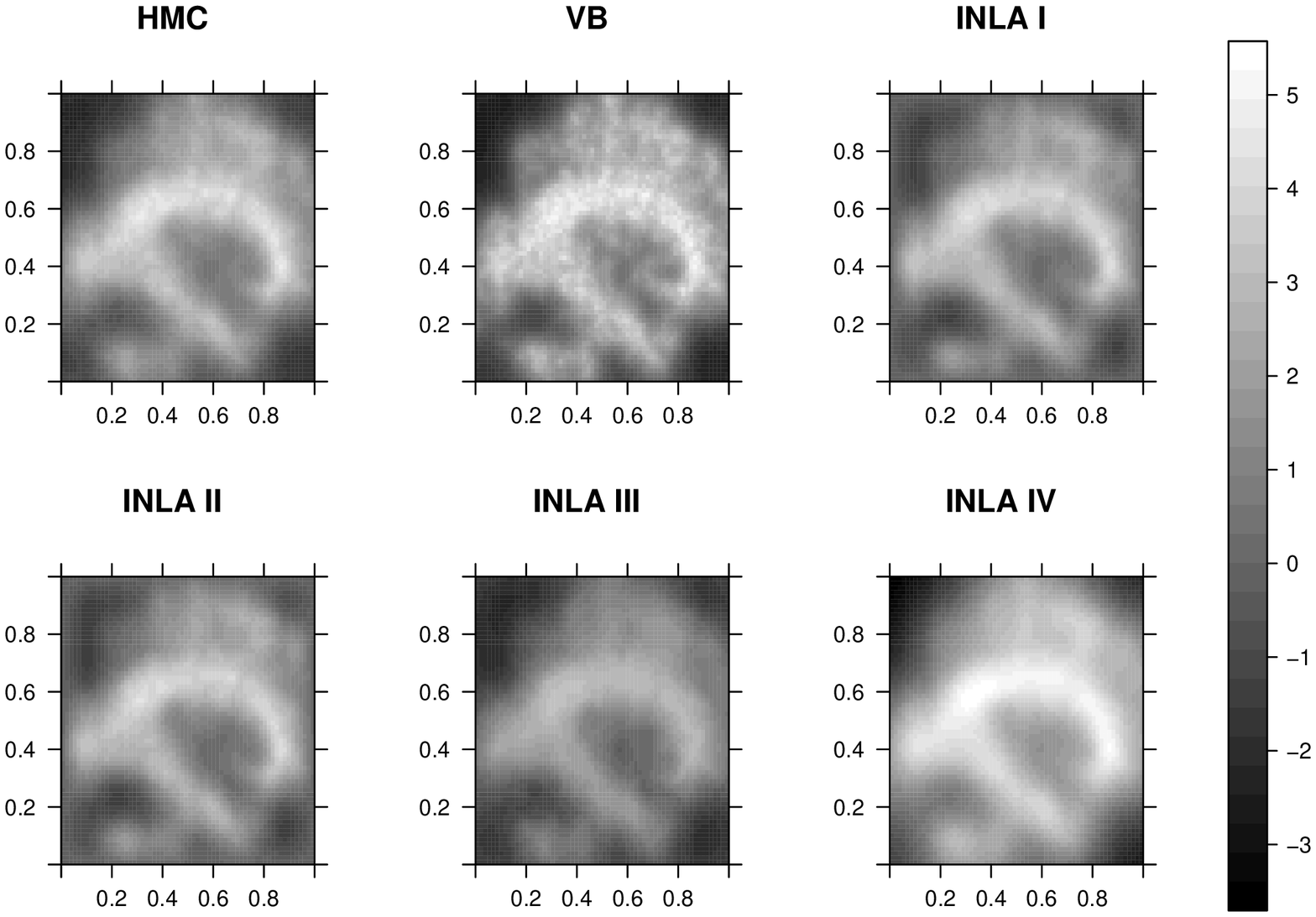}
\caption{Posterior mean of the latent GRF, estimated from HMC, VB and INLA I--IV. MS data set.
} \label{fig9}
\end{center}
\end{figure}

Our second application consists of a point pattern depicting the locations of Multiple Sclerosis (MS) lesions obtained from taking a slab of sagittal slices (10mm thick) obtained from magnetic resonance imaging from a cohort of MS patient and converting the spatial domain to the unit square. The point pattern consists of 1950 locations and is depicted in Figure \ref{fig5}(b). Aside from the application this dataset differs from the first in that the observed level of aggregation is higher and the points are more unevenly distributed. The method of minimum contrast is used to select values of
$\hat{\delta}=1.165$ and $\hat{{\nu}}=1$ for the covariance functions. The posterior mean of the log-intensity values are depicted in Figure \ref{fig9}. In this case all methods seem to capture the same general features of the image.

\begin{figure}[!ht]
\begin{center}
\includegraphics[width=\linewidth]{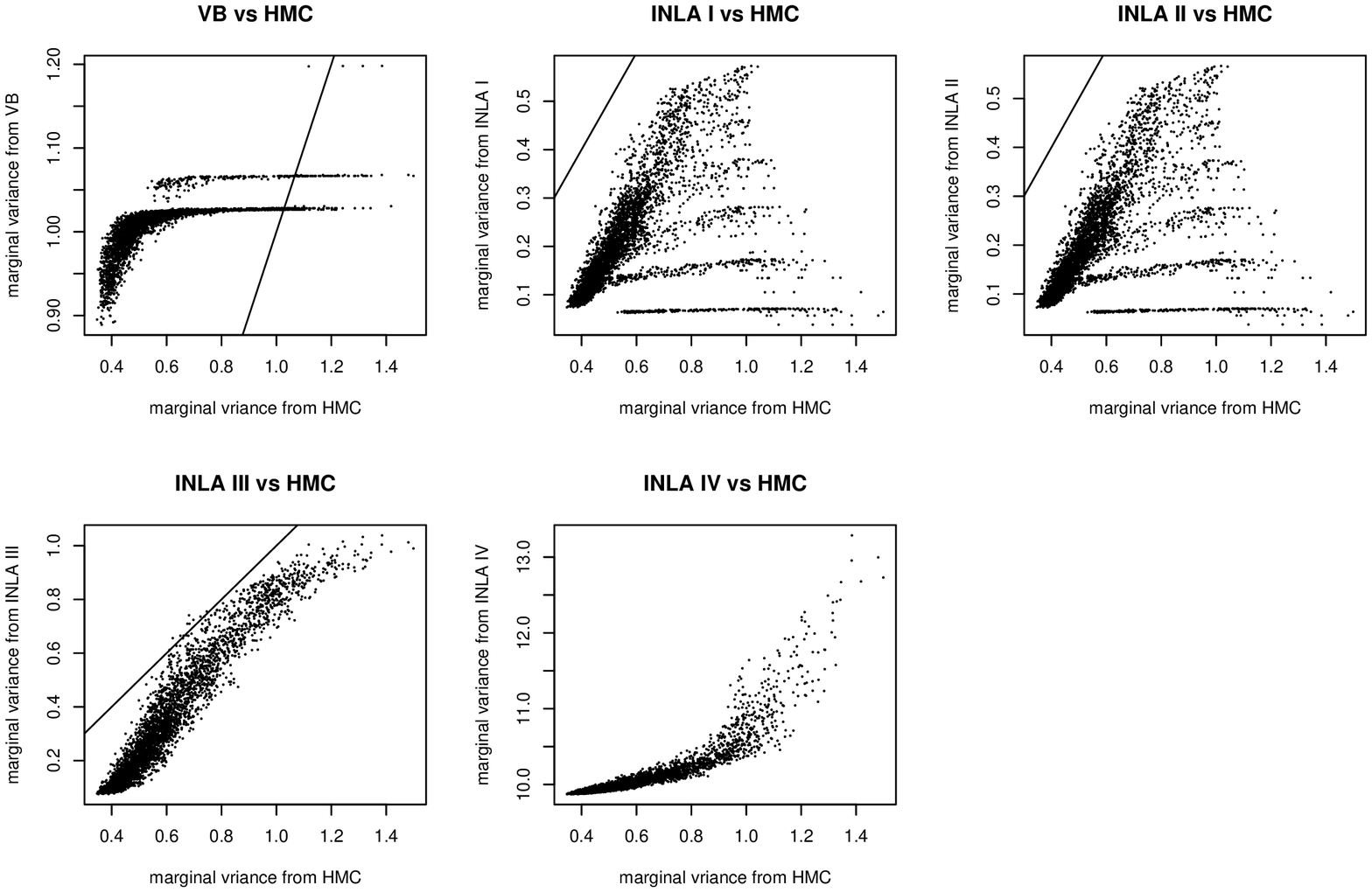}
\caption{Scatter plot of the marginal posterior variance of the latent GRF from VB  and INLA I-IV compared with those from HMC for the MS data set.
} \label{fig10}
\end{center}
\end{figure}

Table \ref{varreal2} presents posterior summaries of the hyper-parameters. The point estimates for $\mu$ obtained from all approaches are fairly consistent while, relative to HMC, posterior variance is not well estimated by either VB or INLA I-IV. Relative to HMC the precision $\sigma^{-2}$ is not well estimated by any of the methods and similarly for the posterior variance of this parameter. As $d_{0.5}$ is estimated rather precisely by HMC, all of VB and INLA I-III either severely under-estimate or over-estimate the posterior variability though this is not the case for INLA IV. Figure \ref{fig10} compares the marginal posterior variance for each cell of the discretized latent field obtained from all of the methods with the posterior variance obtained from HMC. Interestingly, in this case we find that VB over-estimates the posterior variability while INLA I-III all under-estimate the posterior variability to different degrees relative to HMC. The posterior variability arising from INLA IV gives extremely large values (up to 10 times larger than those obtained from HMC) which likely indicates a numerical problem, though we note again that INLA IV does give an adequate representation of the posterior mean for this dataset.

\begin{table}[!ht]
\begin{center}\footnotesize
\begin{tabular}{llr|rrrrr}
\hline \hline
\multicolumn{1}{c}{Parm}	&	\multicolumn{1}{c}{Measure}	&	\multicolumn{1}{c}{}          &\multicolumn{5}{c}{Relative Measure}		\\
	&		&	\multicolumn{1}{c}{HMC}	&	\multicolumn{1}{c}{VB}	&	\multicolumn{1}{c}{INLA I}	&	\multicolumn{1}{c}{INLA II}	&	\multicolumn{1}{c}{INLA III	}&	\multicolumn{1}{c}{INLA IV}	\\ \hline
$\mu$	&	Post. Mean	&	5.098	&	0.959	&	1.074	&	1.072	&	1.193	&	0.814	\\
	&	Post. Var	&	0.301	&	2.638	&	0.081	&	0.084	&	0.197	&	32.731	\\ \hline
$\sigma^{-2}$	&	Post. Mean	&	0.468	&	0.282	&	0.1	&	0.1	&	1.729	&	0.192	\\
	&	Post. Var	&	0.005	&	0.002	&	31.83	&	31.83	&	1.19E-36	&	0.404	\\ \hline
$d_{0.5}$	&	Post. Mean	&	0.18	&	0.886	&	14.379	&	14.379	&	0.503	&	2.789	\\
	&	Post. Var	&	1.30E-04	&	2.41E-10	&	31653.155	&	31653.155	&	2.91E-37	&	0.959	\\ \hline \hline
\end{tabular}
\end{center}
\caption{Summary of parameter estimation for the MS data set. VB and INLA I-IV are relative to HMC.}
\label{varreal2}
\end{table}

\begin{figure}[!ht]
\begin{center}
\begin{subfigure}{0.41\textwidth}
\includegraphics[width=\linewidth]{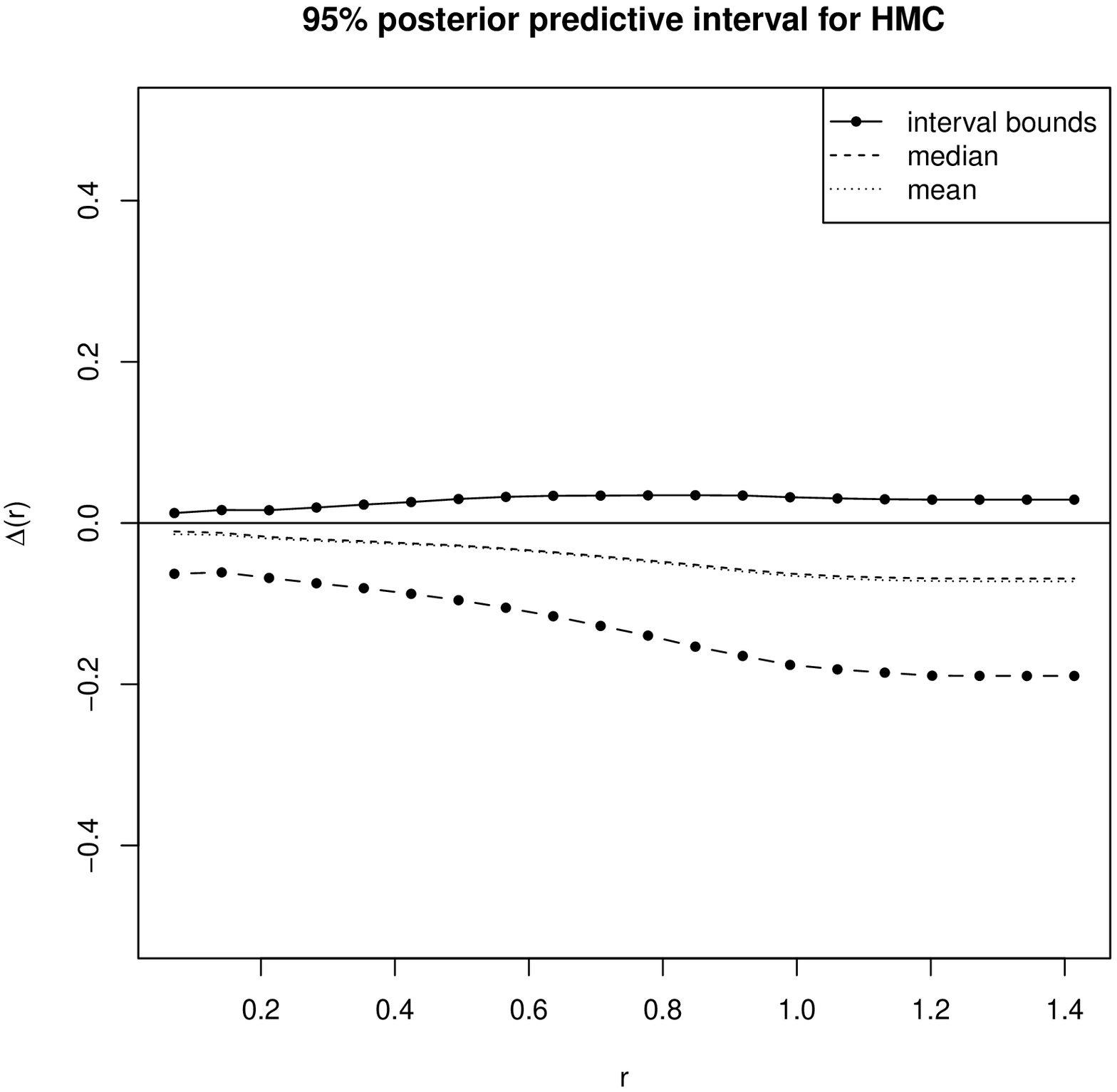}
\caption{} \label{fig11a}
\end{subfigure}
\begin{subfigure}{0.41\textwidth}
\includegraphics[width=\linewidth]{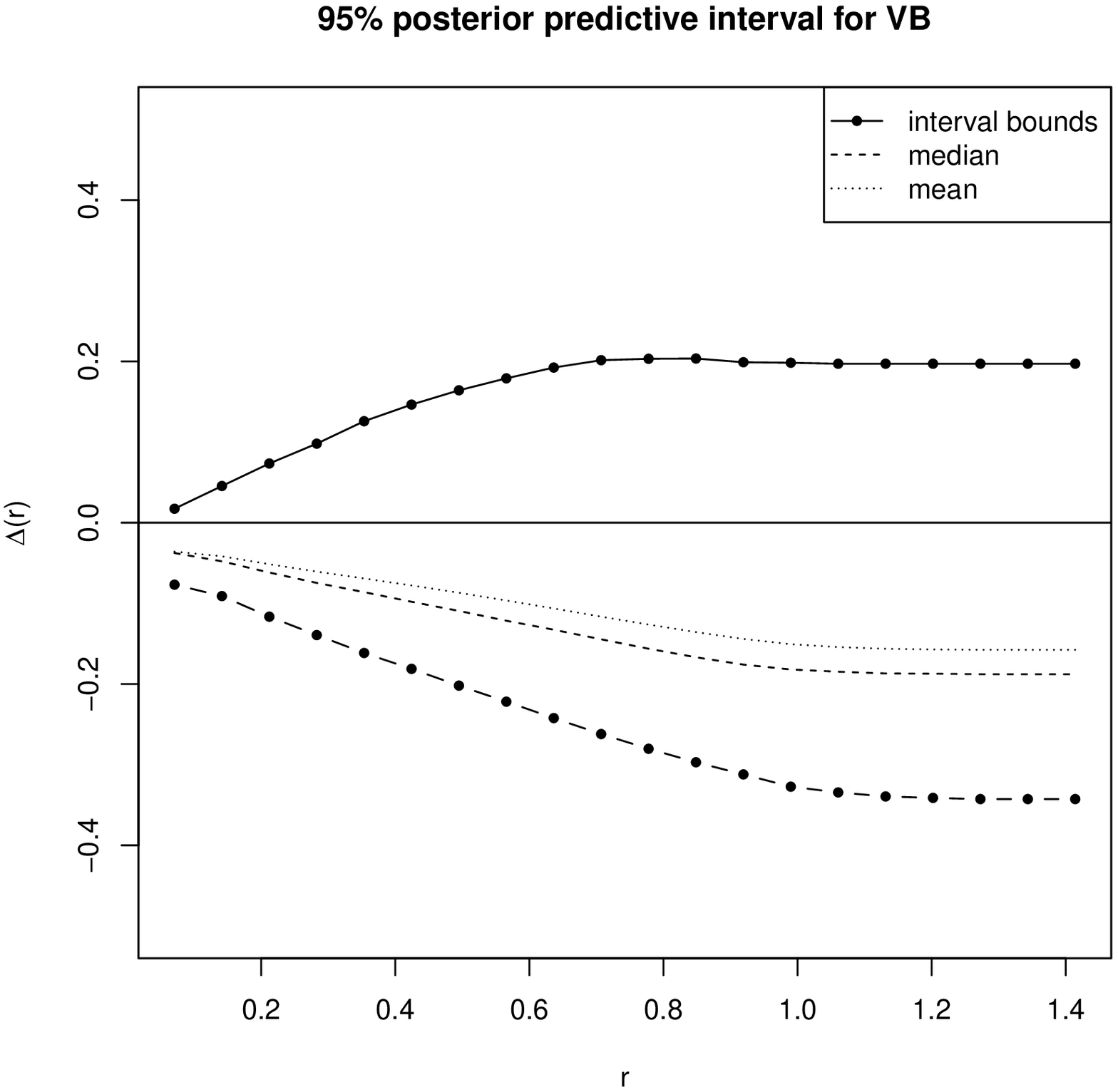}
\caption{} \label{fig11b}
\end{subfigure}
\caption{95\% posterior predictive interval for HMC (a) and VB (b) for the MS data set. The bounds are denoted by solid lines while the mean and median are denoted by dashed lines. These are obtained at 20 distinct distances.} 
\label{fig11}
\end{center}
\end{figure}

Turning to posterior predictive checks which are depicted in Figure \ref{fig11} we see that VB has much wider 95\% posterior predictive intervals than HMC. Although neither algorithm shows a lack of fit, using HMC, the model appears to fit the data better as the mean and median are much closer to zero for all ranges of $r$ and the 95\% predictive intervals are much tighter at each value of $r$.
The greater posterior predictive variability arising from VB may be in part a result of the posterior variance of $\mu$ being over-estimated by VB. In terms of timing, HMC takes 1456s with 2000 iterations and 1000 burn-in. Again, VB actually requires more computation time 1608s with 2763 iterations required for convergence. INLA I takes 47s, INLA II takes 166s, INLA III takes 57s, INLA IV takes 384s. 

\section{Discussion}
We have compared HMC incorporating FFT matrix methods on an extended grid, VB incorporating the Laplace method, and four versions of INLA for Bayesian computation associated with the LGCP model. A number of settings for both simulated and real data have been adopted for these comparisons. Overall, in terms of point estimation of the latent field we do observe some differences in some settings; however, generally, all of HMC, VB, INLA I, and INLA II perform reasonably well, while INLA with SPDE has a tendency to over-smooth the field, though this tendency is reduced as the size of the underlying mesh is increased. Thus if point estimation of the log-intensity is the only objective we recommend the use of INLA I based on the required computation time. If, in addition, inference on hyper-parameters is of importance then it seems clear that HMC and the additional required computation time is necessary. As expected from the literature VB has a tendency to under-estimate posterior variability although on occasion it also seems to over-estimate posterior variability (see e.g. \cite{daunizeau2009variational} for discussion of the latter issue and how it may also arise with VB). We also find that posterior predictive checking based on VB may not be representative of the true posterior predictive distribution. While VB has been applied successfully in a wide range of applications and is often the method of choice in machine learning the required computation time for the LGCP model and for the settings considered here suggest that it is not as accurate as HMC and not as computationally efficient as INLA I. Of course, our implementation of VB did not incorporate the FFT methods for matrix multiplication as this is not straightforward to implement within the VB framework. One approach that may be worth considering is the use of fixed-form multivariate Gaussian variational approximations which may have improved performance over mean field approximations.

To compare within the four flavors of INLA, the INLA with simplified Laplace method and the full Laplace method are quite similar in terms of accuracy, for the settings considered here; whereas, we see genuine gains in computation with the use of the simplified Laplace version of INLA. In addition to a tendency to exhibit over-smoothing, we have also found that INLA with SPDE can be numerically unstable in some situations and an inappropriate choice of step size in the Newton-Raphson algorithm can lead to convergence problems. Thus the choice of mesh is an important consideration.

%\section*{References}
\bibliographystyle{elsarticle-harv}
\bibliography{article}

\newpage
\section*{\large Appendices}
\section*{Appendix A: Gradient derivation for $\rho$}

For the circulant matrix $\mathbf E$ with base $\mathbf e=(e_0,...,e_{m^2-1})$, the $i^{th}$ eigenvalue is given by:
\begin{equation}
\lambda_i=\sum_{j=0}^{m^2-1} \mathbf e_j \exp(\iota 2 \pi j i / m^2)
\end{equation}
where $\iota=\sqrt{-1}$. For the power exponential family of correlations $e_j=\exp(-\rho d_j^{\delta})$ where $d_j$ is the distance from origin. So we have

\begin{equation}
\lambda_i=\sum_{j=0}^{m^2-1} \exp (-\rho d_j^{\delta}) \exp(\iota 2 \pi j i/ m^2)
\end{equation}

Thus, we have $\mathbf E=\mathbf F \Lambda \mathbf F^H$ where $\mathbf F$ is the matrix of eigenvectors, and $\Lambda$ is a diagonal matrix of eigenvalues with $i^{th}$ value to be $\lambda_i$.

To derive the partial derivative of $\log \pi(\rho \mid \cdot)$, we first derive the partial derivative of $\mathbf E^{\frac12} \gamma$

\begin{eqnarray}
\frac{\partial}{\partial \rho} (\mathbf E^{\frac12} \gamma)_i = (\mathbf F \frac{\partial }{\partial \rho} \Lambda^{\frac12} \mathbf F^H \gamma)_i
\end{eqnarray} 

So we need the partial derivative of each diagonal element of $\Lambda^{1/2}$ w.r.t $\rho$.

\begin{eqnarray}
\frac{\partial \lambda_i^{\frac12}}{\partial \rho} 
&&=\frac{\partial}{\partial \rho} \Big ( \sum_{j=0}^{m^2-1} \exp(-\rho d_j^{\delta}) \exp(\iota 2 \pi j i/m^2)   \Big )^{\frac12} \\
&&= -\frac{1}{2} \lambda_i^{-\frac12} \sum_{j=0}^{m^2-1} d_j^{\delta} e_j \exp(\iota 2 \pi j i/m^2)
\end{eqnarray}

The summand in the last line turns out to be the base of a particular matrix with base $\mathbf e^*=(d_0^{\delta}e_0,...,d_{m^2-1}^{\delta}e_{m^2-1})$. Consider the circulant matrix $\mathbf D$ with base $\mathbf d=(d_j^{\delta},...,d_{m^2-1}^{\delta})$. Then it is easy to show that $\mathbf E^* = \mathbf D \odot \mathbf E$ is a circulant matrix with base $\mathbf e^*=\mathbf d \odot \mathbf e$ where $\odot$ represents element wise multiplication. And $\sum_{j=0}^{m^2-1} d_j^{\delta} e_j \exp(\iota 2 \pi j i/m^2)$ is the $i^{th}$ eigenvalue of $\mathbf E^*$. Call it $\psi_i$. Thus,

\begin{equation}
\frac{\partial \lambda_i^{\frac12}}{\partial \rho}=-\frac12 \lambda_i^{-\frac12}\psi_i
\end{equation}

Putting this all together we have 
\begin{eqnarray}
\frac{\partial}{\partial \rho} (\mathbf E^{\frac12} \gamma)_i 
&&= \frac{\partial}{\partial \rho} -\frac12 (\mathbf F \Lambda^{-\frac12} \Psi \mathbf F^H \gamma)_i \\
&&=-\frac12(\mathbf F \Lambda^{-\frac12} \mathbf F^H \mathbf F \Psi \mathbf F^H \gamma)_i \\
&&=-\frac12 (\mathbf E^{-\frac12} \mathbf E^* \gamma)_i
\end{eqnarray}

Now we can derive the gradient for $\rho$:
\begin{eqnarray}
&&\frac{\partial}{\partial \rho} \log \pi(\rho \mid \cdot) = \frac{\partial}{\partial \rho} \Big \{
\sum_i \big [ y_im_i-A\exp(y_i) \big ]+\log \pi(\rho)  \Big \} \\
&&= \sum_i \frac{\partial}{\partial \rho}  \Big \{ \mu m_i +\sigma (\mathbf E^{\frac12} \gamma)_i m_i -A\exp \big [ \mu+\sigma(\mathbf E^{\frac12}\gamma)_i \big ] \Big \}  +\frac{ \pi^{\prime}(\rho)}{\pi(\rho)} \\
&&=\sum_i \Big \{ \sigma \frac{\partial}{\partial \rho} (\mathbf E^{\frac12} \gamma)_im_i-A \frac{\partial}{\partial \rho} \exp \big [ \mu+\sigma (\mathbf E^{\frac12}\gamma)_i  \big ]   \Big \} \\
&&=-\frac{1}{2} \sigma \sum_i \Big \{ m_i - \frac{A\sigma}{2} \exp \big [ \mu+\sigma(\mathbf E^{\frac12}\gamma)_i \big ] \Big \} (\mathbf E^{-\frac12} \mathbf E^* \gamma)_i  \\
&&=-\frac{\sigma}{2} \Big [ \mathbf m-A\exp \big ( \mu \mathbf 1_{m^2} +\sigma \mathbf E^{\frac12} \gamma \big ) \Big ]^T \mathbf E^{-\frac12} \mathbf E^* \gamma
\end{eqnarray}

where $\mathbf m=(m_1,...,m_{m^2})$ and we use the fact that $\pi^{\prime}(\rho)=0$ for flat prior of $\rho$.

Because $\mathbf E^{\frac12} \mathbf E^* \gamma=\mathbf F \Lambda^{-\frac12} \mathbf F^H \mathbf F \Psi \mathbf F^H \gamma=\mathbf F \Lambda^{-\frac12} \Psi \mathbf F^H \gamma$, we can use the DFT to compute all the matrix operations in the equation above.

\end{document}